\documentclass[twocolumn,preprintnumbers,amsmath,amssymb]{revtex4}

\usepackage{graphicx}
\usepackage{dcolumn}
\usepackage{bm}
\usepackage{siunitx}
\usepackage{xcolor}
\usepackage{placeins}
\usepackage[normalem]{ulem}
\usepackage[colorlinks]{hyperref}
\usepackage{mathrsfs}

\usepackage[T1]{fontenc}

\begin{document}


\title{Magneto-Stark and Zeeman effect as origin of second harmonic generation of excitons in Cu$_2$O}

\author{A. Farenbruch$^{1}$, J. Mund$^{1}$, D. Fr\"ohlich$^{1}$, D. R. Yakovlev$^{1,2}$, M. Bayer$^{1,2}$, M. A. Semina$^{2}$ and M. M. Glazov$^{2}$}

\affiliation{$^{1}$ Experimentelle Physik 2, Technische Universit\"at Dortmund, D-44221 Dortmund, Germany}
\affiliation{$^2$ Ioffe Institute, Russian Academy of Sciences, 194021 St. Petersburg, Russia}

\date{\today}

\begin{abstract}
We report on the experimental and theoretical investigation of magnetic-field-induced second harmonic generation (SHG) and two-photon absorption (TPA) of excited exciton states ($n \geqslant 3$) of the yellow series in the cuprous oxide Cu$_2$O. In this centrosymmetric material, SHG can occur due to constructive interplay of electric dipole and electric quadrupole/magnetic dipole transitions for light propagating along the low-symmetry directions $[111]$ or $[112]$. By application of a magnetic field in Voigt configuration, SHG gets also allowed for excitation along the $[110]$-axis and even the high-symmetry cubic direction $[001]$. Combining a symmetry analysis and a microscopic theory, we uncover the two key contributions to the magnetic-field-induced SHG: the Zeeman effect and the magneto-Stark effect. We demonstrate systematic dependencies of the SHG intensity on the linear polarization angles of the ingoing fundamental laser and the outgoing SHG beam, complementary to the manuscript by Rommel et al.~\cite{RommelSHG}. In general, the resulting contour plots in combination with a symmetry analysis allow one to determine uniquely the character of involved transitions. Moreover, we can separate in magnetic field the Zeeman and the magneto-Stark effect through appropriate choice of the experimental geometry and polarization configuration. We present a microscopic theory of the second harmonic generation of excitons in a centrosymmetric cubic semiconductor taking into account the symmetry and the band structure of cuprous oxide. Based on the developed microscopic theory we identify the main contributions to the second-order nonlinear susceptibility of $S$-, $P$- and $D$-excitons. We analyze the redistribution of SHG intensities between the excitonic states both in the absence and presence of the magnetic field and show good agreement with the experimental data. With increasing exciton principal quantum number the magneto-Stark effect overpowers the influence of the Zeeman effect.
\end{abstract}

\maketitle

\section{INTRODUCTION}
\label{introduction}
Nonlinear optical experiments, including multi-photon absorption, higher harmonics generation, multiple wave-mixing, etc., involve more than one photon in the elementary excitation or emission process. These methods form a well-established spectroscopic toolbox for the investigation of electronic properties, which in many cases are not accessible to linear optical experiments such as one-photon absorption or linear reflectivity~\cite{Shen,Boyd,GP}. Nonlinear optical spectroscopy has turned out to be particularly valuable for studying semiconductors ~\cite{Pavlov05PRL,Pisarev10,glazov:review,Glazov2017,semina:18,Mund1,Mund2} whose optical properties are largely controlled by excitons, hydrogen-like bound states of electrons and holes~\cite{excitons:RS,Klingshirn}.

Importantly, in this respect, different excitonic states can be active in linear and nonlinear optical processes, underlining the complementarity of these techniques. This is particularly prominent in the centrosymmetric semiconductor cuprous oxide Cu$_2$O, in which the Mott-Wannier excitons were discovered~\cite{GK}: odd-parity $P$-shell excitons are mainly active in linear optical absorption, while even parity $S$-shell excitons provide the key contribution to two-photon absorption~\cite{Uihlein}. The combination of these specific selection rules with the high quality of natural Cu$_2$O crystals has enabled demonstration of the Rydberg series of well-resolved $P$-excitons up to the principal quantum number~$n=25$~\cite{Nature}, and of $S$- and $D$-excitons up to $n = 5$~\cite{Uihlein}. The large, up to micrometer, radii of highly excited Rydberg excitons make them quite susceptible to external electric and magnetic fields~\cite{Heck,chaos} and also enhance the optical nonlinearities, e.g., due to Rydberg and plasma blockade effects~\cite{Nature,plasma}, so far studied by linear spectroscopy.

The nonlinear optical properties of Cu$_2$O with its prominent excitonic features have been attracting researchers' attention already early on~\cite{Uihlein,Froehlich,Shen:Cu2O}. The continuous development of optical spectroscopy techniques has recently made it possible to observe second harmonic generation (SHG) in Cu$_2$O crystals with high spectral resolution, despite of the broadband excitation with short light pulses~\cite{Mund1,Mund2}. Being forbidden in the electric dipole approximation, SHG arises due to suitable combinations of electric dipole, electric quadrupole and magnetic dipole transitions. In that way it was possible to extend the series of observed $S$-excitons up to $n = 9$ and also resolve $D$-excitons up to $n = 7$~\cite{Mund1}.

Although an external magnetic field does not break the space inversion ($\mathcal P$) symmetry, it results in a non-trivial state mixing and, through the time-reversal ($\mathcal T$) symmetry breaking, quantum chaotic behavior may arise for Rydberg excitons~\cite{chaos}, see also Refs.~\cite{Zhilich1969,Schweiner_1,Rommel} for a review of the linear magneto-optics in Cu$_2$O. As was shown in Ref.~\cite{Mund1}, SHG on the low energy excitons in Cu$_2$O emerges in a magnetic field even along high symmetry directions, resulting in rich spectra consisting of multiple lines. Reference~\cite{Mund2} reported SHG on the $1S$ exciton in Cu$_2$O at zero field in a symmetry-forbidden geometry (see also Ref.~\cite{Shen:Cu2O}) as a consequence of sample-inherent strain, breaking the symmetry and activating nominally forbidden excitons for SHG. Interestingly, SHG was demonstrated to be an extremely sensitive strain sensor on a level of parts in a million.

A systematic experimental and theoretical study of magnetic-field-induced SHG of excited excitons of Cu$_2$O is, however, lacking. The aim of this paper is to close this gap. In detail, we present a nonlinear magneto-optical study of higher lying excitons $(n\geqslant 3)$ for different crystalline orientations in magnetic fields up to $\SI{10}{\tesla}$. On the experimental side, we mainly focus on the SHG forbidden directions (e.g., when the light propagates along a $[110]$ crystalline axis), where SHG is not allowed in the absence of a magnetic field. Such measurements are of special interest, since SHG becomes allowed in the presence of the field by the Zeeman effect (ZE) or the magneto-Stark effect (MSE). The latter effect, namely, originates from the mixing of odd- and even-parity excitons due to the equivalent electric field arising from a magnetic field normal to the direction of exciton motion. The MSE demonstration is particularly important since it directly evidences exciton motion in the crystal~\cite{Hopfield1,Hopfield2,konstantinov}. The MSE was first observed in one-photon absorption on the $1S$ resonance in Ref.~\cite{Gross} and recently on the yellow exciton series of Cu$_2$O in Ref.~\cite{Rommel}. The MSE also controls SHG on the excitons in the noncentrosymmetric semiconductor ZnO~\cite{Lafrentz}. Thus it is interesting to assess this effect in SHG also in centrosymmetric crystals.

The SHG effect in cubic noncentrosymmetric crystals has been extensively studied in literature and the associated symmetry analysis is a textbook problem~\cite{Shen,Boyd}. Cuprous oxide has a centrosymmetric structure where SHG is forbidden if the effects of the radiation wavevector (spatial dispersion) and external fields are disregarded. The analysis of the interplay of the wavevector related and the magnetic field induced effects becomes already nontrivial on the phenomenological level. Also, the identification of the microscopic pathways of SHG and evaluation of the contributions of each contributing mechanism to the second-order nonlinear susceptibility has, to our best knowledge, not been addressed in the literature. Thus, on the theoretical side we combine a symmetry-based phenomenological analysis of TPA and SHG in Cu$_2$O with a microscopic theory which demonstrates the main underlying mechanisms both in absence and presence of a magnetic field. Particularly, in the framework of the symmetry-based approach we present full rotation anisotropies of SHG for arbitrary polarizations for the incident $\mathbf E^\omega$ and outgoing $\mathbf E^{2\omega}$ light fields relative to each other, going beyond Refs.~\cite{Mund1,Mund2} where only two distinct geometries ($\mathbf E^{\omega} \parallel \mathbf E^{2\omega}$ and $\mathbf E^{\omega} \perp \mathbf E^{2\omega}$) were investigated. This analysis allows us to select the most appropriate experimental setting to observe and distinguish different mechanisms of SHG. On the microscopic level we identify the main pathways for the SHG process in the centrosymmetric crystal and present general expressions for the second-order susceptibility. For particular excitonic states, e.g., the $S$-shell, $P$-shell, and $D$-shell states (using atomic nomenclature where $S,P,D,\ldots$ denote the orbital angular momentum of the exciton envelope function) we present simplified expressions for the susceptibility which allows a direct comparison of the relative SHG contributions of the different states. The Zeeman and the magneto-Stark effects are analyzed in detail. We demonstrate that while in the absence of a magnetic field the odd $P$-shell excitons provide parametrically small contributions to the SHG as compared to the $S$-shell excitons, the MSE can result in equally strong SHG on the $S$- and $P$-shell states. 

We study theoretically SHG also on the $D$-shell excitons (with $\Gamma_1^+$ and $\Gamma_3^+$ symmetry in the notations of Ref.~\cite{Koster}) that are not coupled to the $S$-shell states. The main predictions are confirmed by the experimental data. In agreement with the model we observe the strongest SHG on $S/D$-mixed states ($\Gamma_5^+$ symmetry according to Ref.~\cite{Koster}), while much weaker SHG signals are found on the $D$-excitons that are disjunct from the $S$-states. Nevertheless, also these states can be clearly identified in SHG through their distinct polarization dependence allowing to separate them from the dominant processes. In this work we use a combination of symmetry analysis and perturbation theory to study the effect of magnetic field on the SHG and TPA processes. Microscopic calculations of the excitonic states in a magnetic field for fulfilling the conditions of SHG and TPA are presented in the counterpart manuscript~\cite{RommelSHG}.

The paper is organized as follows: Sec.~\ref{shg polarization dependences} presents the phenomenological analysis of SHG in Cu$_2$O, based on the coupling coefficients of Ref.~\cite{Koster} for the derivation of polarization dependences in different crystalline and magnetic field orientation configurations as well as different scenarios of excitation (electric dipole and quadrupole as well as magnetic dipole) and magnetic-field-induced effects (ZE and MSE). This analysis allows us to identify the main SHG mechanisms due to the symmetry of the perturbations. Further, in Sec.~\ref{sec:micro} the microscopic theory is presented from which the relative importance of the SHG processes in Cu$_2$O is assessed. Sec.~\ref{sec:experiment} describes the samples and the experimental technique, the experimental results are given in Sec.~\ref{sec:experimental results} where they are also set in relation with the models in the preceding sections. The paper is summarized by a brief conclusion and an outlook.

\section{SHG polarization dependences}
\label{shg polarization dependences}

\subsection{Phenomenological analysis}
\label{subsec:simplified}

The point symmetry of the system imposes restrictions on the linear and nonlinear optical processes and allows us to determine the basic geometry and polarization dependences of SHG without resorting to a microscopic model. Furthermore, the symmetry analysis makes it possible to establish signatures of particular excitonic states in the SHG spectra, from which the involved types of transitions can be derived. In this section we perform a phenomenological analysis of SHG in Cu$_2$O, while the microscopic model of SHG is presented in Sec.~\ref{sec:micro}.

We recall that Cu$_2$O is described by the $O_h$ point symmetry group which includes spatial inversion. Thus, SHG is allowed only with taking into account the light wave vector $\bf k$ or the magnetic field of the electromagnetic wave (this is mathematically the same as the alternating magnetic field in the wave $\tilde{\bf B}\propto [\bf k\times\bf E]$). Phenomenologically, in the absence of an external magnetic field, SHG in Cu$_2$O is described by the following relation 
\begin{equation}
	\label{SHG:cryst}
	P_i=\chi_{ijlm}k_jE_lE_m,
\end{equation}
where $P_i$ is the induced polarization component at twice the frequency of the incident light, $\chi_{ijlm}$ are the susceptibility tensor components and $E_k$ are the components of the electric field of the light at the fundamental frequency, $i,j,l,m$ denote the Cartesian components. The tensor $\chi_{ijlm}$ is symmetric with respect to permutation of the two last subscripts; summation over repeated subscripts is assumed. The process of SHG can be understood as a two-photon excitation followed by coherent single photon emission at double frequency.

The number of independent components of the susceptibility can be readily found from the symmetry analysis. Due to the permutation symmetry of $\chi_{ijlm}$ only the symmetrized products $\{E_m E_l\}_{\rm sym}$ are relevant, they transform according to the reducible representation 
\[
\{E_lE_m\}_{\rm sym} \sim \mathscr D_{EE}=\Gamma_1^+ + \Gamma_3^+ + \Gamma_5^+
\]
of the $O_h$ point group. The wave vector components, on the other hand, transform according to $\Gamma_4^-$. Since
\[
\Gamma_4^- \times \mathscr D_{EE} = \Gamma_2^- + \Gamma_3^- + 3\Gamma_4^- + 2\Gamma_5^-,
\]
there are three contributions to the crystallographic SHG. However, two of those are $\mathbf P \propto \mathbf k E^2$ and $\bf P \propto (\bf k \cdot \bf E)\bf E$ which can be disregarded for transverse fields, since in the first case the polarization is longitudinal and in the second case the scalar product $\bf k\cdot \bf E$ vanishes. As a result, there is only one independent constant $\chi_c$ and SHG is described by the phenomenological relation
\begin{equation}
	\label{SHG:cryst:1}
	P_i = \chi_c k_i (2E_i^2 - E_{i+1}^2 - E_{i-1}^2),
\end{equation}
where $i=x,y,z$ is the Cartesian index. We use the cyclic rule in this notation, e.g., for $i=z$ we have $i+1=x$ and $i-1=y=i+2$. Here and in what follows we use the cubic axes with ${x}\parallel [100]$, ${y}\parallel [010]$ and ${z}\parallel [001]$.

Let us now turn to the magnetic-field-induced SHG. In the linear field regime one has the following phenomenological relation:
\begin{equation}
	\label{SHG:B}
	P_i(2\omega)=\chi_{ijlmn}k_j B_l E_mE_n.
\end{equation}
The product $k_j B_l$ can be recast into symmetrized and antisymmetrized parts. The antisymmetrized part corresponds to the vector product $[\bf k\times \bf B]$, it transforms as a vector, i.e., according to the $\Gamma_4^-$ irreducible representation. The contributions containing such an asymmetric product can be attributed to the magneto-Stark effect because $[\bf k\times \bf B]$ de facto acts as an electric field which mixes active and inactive exciton states, see details below. Accordingly, we find that the magneto-Stark contribution is described by three independent constants $\chi^{MS}_{1\ldots 3}$ and
\begin{subequations}
	\label{SHG:MS}
	\begin{align}
		&\mathbf P = \chi_1^{MS} [\bf k \times \bf B] E^2,\\
		&\mathbf P = \chi_2^{MS} \bf E ([\bf k \times \bf B] \bf E),\\
		&P_i = \chi_3^{MS} [\mathbf k \times \mathbf B]_i(2E_i^2 - E_{i+1}^2 - E_{i-1}^2).
	\end{align}
\end{subequations}
The remaining contributions arise from the symmetrized products $\{k_jB_k\}_{\rm sym}$ which transform according to the reducible representation 
\[
\{k_jB_l\}_{\rm sym} \sim \mathscr D_{qB}=\Gamma_1^- + \Gamma_3^- + \Gamma_5^-.
\]
These contributions can be tentatively assigned to the Zeeman effect of the magnetic field.
The product 
\[
\mathscr D_{qB} \times \mathscr D_{EE} = 3\Gamma_4^-+\ldots,
\]
where the dots denote omitted contributions which transform according to other irreducible representations. Thus, there are three contributions, one of which, $\bf P \propto (\bf k\bf E) [\bf E \times \bf B]$, vanishes for transversal fields. The remaining two contributions take the form
\begin{subequations}
	\label{SHG:Z}
	\begin{align}
		&P_i = \chi_1^Z \{k_{i+1} B_{i-1}\}_{\rm sym} (E_{i+1}^2 - E_{i-1}^2),\\
		& P_i = \chi_2^Z \sum_\pm \{k_{i\mp 1} B_i\}_{\rm sym}\{E_{i\pm 1} E_i\}_{\rm sym}.
	\end{align}
\end{subequations}
The phenomenological equations~\eqref{SHG:MS} and \eqref{SHG:Z} describe the magnetic-field-induced SHG in Cu$_2$O. Our next step is to identify the symmetries of the excitonic states in Cu$_2$O and analyze their contributions to SHG.

\subsection{Band structure and symmetry of excitonic states}
\label{subsec:bands}

\begin{figure}[h]
	\includegraphics[width=0.65\linewidth]{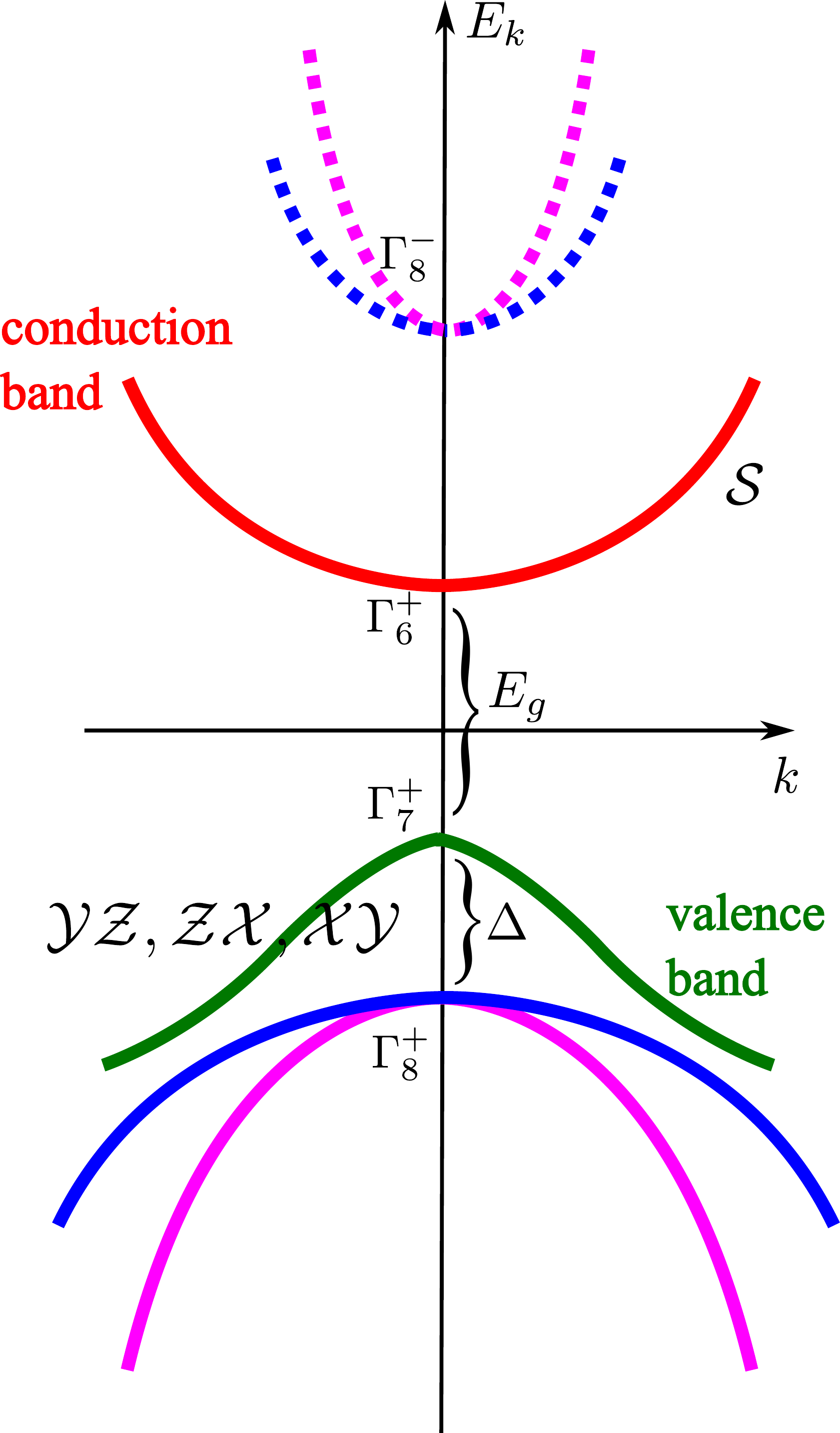}
	\caption{Schematic illustration of the band structure in Cu$_2$O. The lowest conduction band ($\Gamma_6^+$) and topmost valence bands ($\Gamma_7^+$ and $\Gamma_8^+$) are marked by solid lines. The odd parity excited conduction bands ($\Gamma_8^-$) are shown by dotted lines. The bands are labeled by the corresponding irreducible representations of the $O_h$ point symmetry group. The types of underlying orbital Bloch functions are also indicated at the bands.}\label{fig:bands}
\end{figure}

Figure~\ref{fig:bands} illustrates the band diagram of Cu$_2$O. In the center of the Brillouin zone, the top valence band states are formed from the $\Gamma_5^+$ orbital functions which transform as $\mathcal Y\mathcal Z$, $\mathcal Z\mathcal X$ and $\mathcal X\mathcal Y$. The spin-orbit coupling splits the six-fold degenerate valence band into the $\Gamma_7^+$ (two-fold degenerate, topmost) and $\Gamma_8^+$ (four-fold degenerate, bottom) branches. The conduction band has $\Gamma_6^+$ symmetry and the Bloch functions are formed from the products of the invariant $\mathcal S$-type orbitals ($\Gamma_1^+$ symmetry) and the basic spinors ($\Gamma_6^+$)~\cite{Uihlein}. The higher conduction band is formed from odd-parity states transforming according to $\Gamma_3^-$ and, with account for the spin, its Bloch functions form the basic functions of the $\Gamma_8^-$ irreducible representation~\cite{Gross}. Note that below the $\Gamma_8^+$ bands there are a doublet and a quadruplet of $\Gamma_7^-$ and $\Gamma_8^-$ bands arising from the orbital functions $\mathcal X$, $\mathcal Y$ and $\mathcal Z$ (not shown)~\cite{PhysRevB.21.1549,PhysRevB.56.7189}.

The parity of the conduction and valence bands is the same, therefore, optical transitions take place with account for the $\bf k \cdot \bf p$-mixing with the remote odd parity bands ($\mathbf k$ is the electron wavevector and $\mathbf p$ is the interband momentum operator). Within the minimal model we can assume that the transitions take place with admixture of the upper conduction band of $\Gamma_8^-$ symmetry since both direct products $\Gamma_7^+\times \Gamma_8^-$ and $\Gamma_6^+\times \Gamma_8^-$ contain the irreducible representation $\Gamma_4^-$ according to which the components of a vector transform. The actual situation is more difficult because in the absence of the spin-orbit coupling dipole transitions between the $\Gamma_1^+$ and $\Gamma_3^-$ bands are forbidden. The bottom valence band $\Gamma_8^-$ (involving $\Gamma_4^-$ orbital states) can play the role of the intermediate state in multi-photon processes.

The symmetry of the excitonic state is described by the product of the irreducible representations for the hole state in the valence band, $\Gamma_7^+$, the electron state in the conduction band, $\Gamma_6^+$, and that of the envelope function $\mathscr D_{env}$. We will be mainly interested in $S$-shell, $D$-shell and also $P$-shell excitons. For $S$-excitons, $\mathscr D_{env}=\Gamma_1^+$, so that they transform according to either $\Gamma_2^+$ (paraexciton) or the three-fold degenerate $\Gamma_5^+$ (orthoexciton). For $D$-shell excitons $\mathscr D_{env}=\Gamma_3^+$ or $\Gamma_5^+$, resulting in $\Gamma_5^+$ states (which are efficiently mixed with $S$-excitons due to the complex valence band structure), as well as in $\Gamma_1^+$, $\Gamma_3^+$ and $\Gamma_4^+$ states which are not mixed with the $S$-excitons. Finally, the $P$-excitons give rise to a variety of symmetries of states out of which we will be interested in those transforming according to $\Gamma_4^-$, i.e. those, which are optically active in one-photon processes in the dipole approximation.

Knowledge of the exciton state symmetry allows one to determine the selection rules for the excitation and emission processes and, finally, the polarization dependences for the TPA and SHG. Since in emission the $S$- and $D$-excitons require an electric quadrupole (or magnetic dipole) process, we mainly focus on the states which can be directly excited by two photons, these are the states of $\Gamma_1^+$, $\Gamma_3^+$ and $\Gamma_5^+$ symmetry.

In Ref.~\cite{Mund1} we considered only the $\Gamma_5^+$ contributions which get allowed by their admixtures to the $\Gamma_5^+$ $S$-excitons. Since angular momentum is no longer a good quantum number, the other $D$-excitons can also lead to a SHG signal, which might, however, be weaker. In the following we will first derive the polarization dependences for the processes expected to be dominant ($\Gamma_5^+$ symmetry, see Sec.~\ref{sec:dom}) and then for the processes expected to be weaker ($\Gamma_1^+$ and $\Gamma_3^+$ symmetry, see Sec.~\ref{sec:weak}). Results on SHG for the $S$-paraexciton ($\Gamma_2^+$) will be reported elsewhere. In Section~\ref{sec:experimental results} we will also present SHG spectra for the weaker processes, which nevertheless can be clearly distinguished by their polarization dependence from the dominant processes.

\subsection{Dominant Processes}
\label{sec:dom}

In this section we will derive polarization dependences for SHG-allowed as well as SHG-forbidden crystalline orientations in a magnetic field. 

\begin{figure}[h]
	\begin{center}
		\includegraphics[width=0.48\textwidth]{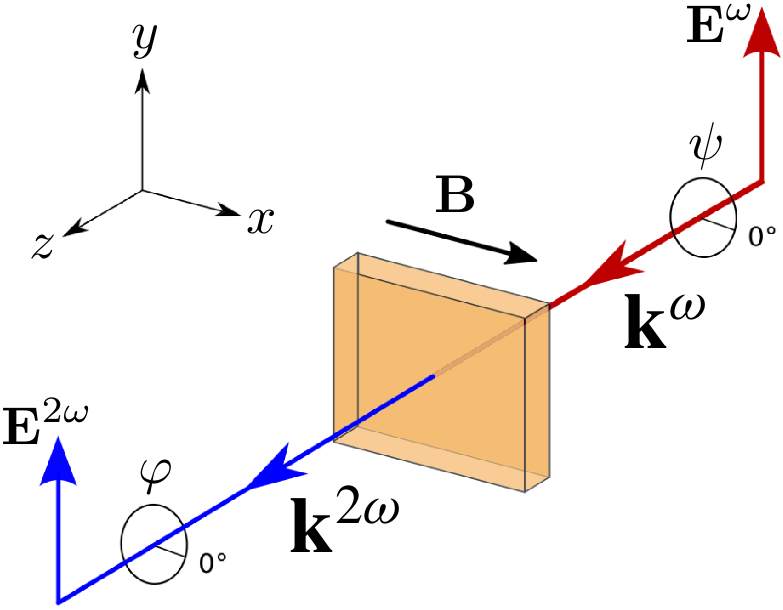}
		\caption[Scan]{Experimental geometry in Voigt configuration with the Cu$_2$O crystal oriented such that $z\parallel \textbf{k}\parallel[1\bar 1 0]$, $x\parallel \mathbf{E}^{\omega,2\omega}(0^\circ)\parallel[110]$ and $y\parallel \mathbf{E}^{\omega,2\omega}(90^\circ)\parallel[001]$.}
		\label{fig:orientation}
	\end{center}
\end{figure}

As was shown allready in the first derivation of two-photon selection rules \cite{Toyozawa} and later for three-photon processes \cite{Pasquarello}, one can separate the transition probability of nonlinear processes into the product of a geometrical part and a dynamical part. From the detailed polarization dependences, which allow us to distinguish different physical mechanisms of excitation, we derive the geometrical part simply by application of group theory, using the tables of irreducible representations and coupling coefficients by Koster, Dimmock, Wheeler and Statz \cite{Koster}. In the dynamical part, however, one has to take into account the specific electronic transitions determined by the band structure, excitonic and polaritonic effects. Excitonic effects are discussed in detail in Sec.~\ref{sec:micro}, while the polaritonic effects can be taken into account following Ref.~\cite{semina:18}, but are negligible for the studied system. The derivation presented below is an extension of the results reported in Ref.~\cite{Mund1} in two aspects. (i) Detailed SHG polarization dependences are derived for experiments in a magnetic field. (ii) We demonstrate here two-dimensional ($2$D) plots (intensity maps vs. the linear polarization angles $\psi$ and $\varphi$ of the ingoing and outgoing photons, respectively), which offer an elegant way to extract polarization dependences in order to distinguish between different mechanisms of SHG in allowed and forbidden crystalline orientations. By contrast, in Ref.~\cite{Mund1} SHG only for the two configurations of parallel and perpendicular polarizations of the ingoing and outgoing light was analyzed.

It was already demonstrated in Ref.~\cite{Mund1} that SHG can be observed in forbidden directions (e.g. $\mathbf k\parallel[001]$ and $[110]$) by applying a magnetic field. In this paper, we will show that by use of group theory \cite{Koster} we can derive polarization dependences for magnetic-field-induced SHG signals. Experiments in Voigt configuration are of special interest, since there are two mechanisms, which lead to SHG signals: (i) the Zeeman effect (ZE) \cite{Schweiner_1}, which is described by the even parity perturbation operator $\Gamma^+_4$ (magnetic field $\mathbf B$) and (ii) the magneto-Stark effect (MSE) \cite{Rommel}, which is taken into account by the odd-parity perturbation operator $\Gamma^-_4$ (effective electric field $\mathbf E_{\text{MSE}} \sim \mathbf k \times \mathbf B $).

\begin{figure}[h]
	\begin{center}
		\includegraphics[width=0.48\textwidth]{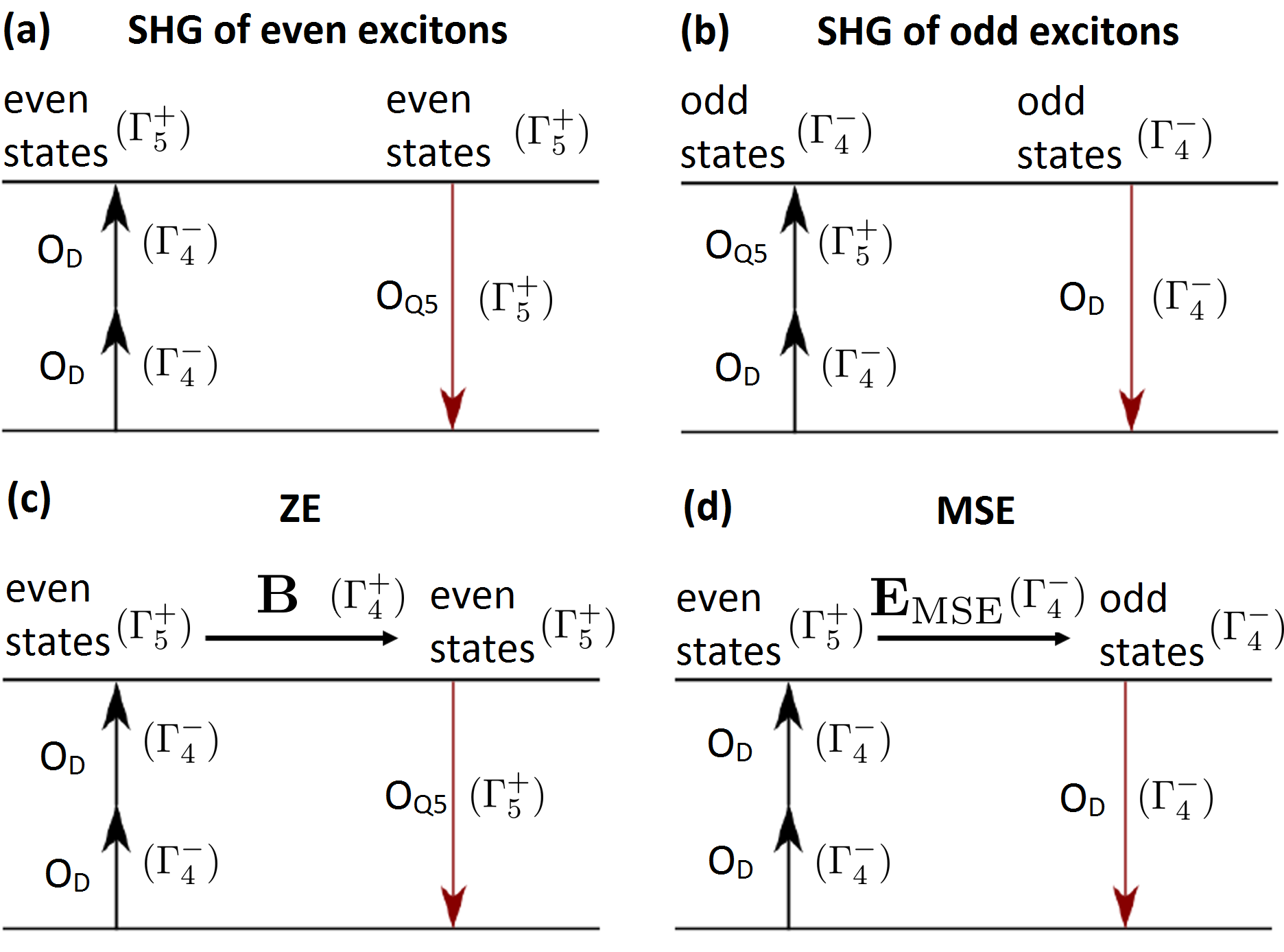}
		\caption[Scan]{Schematics of the SHG process for (a) even- and (b) odd-parity excitons at zero magnetic field and in finite field for (c) the Zeeman effect (ZE) and (d) the magneto-Stark effect (MSE) involving the even parity excitons.}
		\label{fig:scheme}
	\end{center}
\end{figure}

In Figure~\ref{fig:scheme} we sketch the different scenarios of SHG processes. The excitation of even excitonic states of $\Gamma_5^+$ symmetry is possible via two dipole processes (with intermediate states in the remote bands, see Sec.~\ref{sec:micro}). The excitation of odd states, e.g., $P$-shell excitons of $\Gamma_4^-$ symmetry is possible by a combination of a dipole and a quadrupole transition, see Sec.~\ref{sec:micro} for details. The emission of the $\Gamma_5^+$ excitons takes place in the quadropole approximation, while the emission of the $\Gamma_4^-$ states is dipole allowed. Figures \ref{fig:scheme}(a) and \ref{fig:scheme}(b) describe the resulting zero-field case for allowed SHG transitions \cite{Mund1}. For both scenarios as well as for the forbidden directions, the ZE and MSE lead to magnetic-field-induced SHG as depicted in Figs.~\ref{fig:scheme}(c) and \ref{fig:scheme}(d) for even parity excitons. The mixing mechanism, however, is different. For the ZE the mixing takes place with the quadrupolar-allowed even exciton, while the MSE effect results in the admixture of the dipole allowed odd-parity state to the even exciton. The magnetic-field-induced effects on odd parity excitons will be discussed in detail in Sec.~\ref{sec:micro}.

As mentioned above, two-dimensional presentations are very helpful to identify the underlying SHG mechanisms and accordingly select the specific experimental configuration to separate resonances of ZE or MSE origin. For an arbitrary configuration, however, interference between both effects has to be taken into account. It turns out that in the Faraday configuration magnetic-field-induced effects will not appear along the forbidden directions. On the other hand, in the case of allowed SHG transitions (e.g. along the $[111]$ and $[112]$ direction) one expects field-induced effects in addition to SHG in the zero-field case. Thus, in the general case, the three contributions (zero-field SHG, ZE and MSE) interfere in Voigt configuration. It will be shown, that by selecting proper polarization configurations one can distinguish between different terms. 

Following the schematic representations in Fig.~\ref{fig:scheme}, the SHG process can be separated into two steps: (i) Two-photon excitation via combined dipole-dipole or dipole-quadrupole transitions and (ii) one-photon emission via dipole or quadrupole emission processes. Dipole and quadrupole processes for the $\Gamma_5^+$ $S/D$-excitons and $\Gamma^-_4$ $P$-excitons, which are expected to be the dominant contributions, are considered in the following. In Section \ref{sec:weak} weaker contributions from $D$-envelope excitons of $\Gamma^+_1$ and $\Gamma^+_3$ symmetry will be considered. 

The selection rules for the TPA and SHG processes depicted in Fig.~\ref{fig:scheme} can be easily presented in the cubic axes ($x,y,z$). In order to describe the phenomenology of SHG for arbitrary light propagation direction and light polarization we introduce the polarization rotation matrix given in Eq.~\eqref{eqn:mrot}
\begin{widetext}
	\begin{align}
		\mathbf M_{\text{rot}}(\mathbf k, \psi)=
		\begin{pmatrix}
			k_1^2(1-\cos \psi)+\cos\psi & k_1k_2(1-\cos \psi)-k_3\sin\psi & k_1k_3(1-\cos \psi)+k_2\sin\psi \\
			k_2k_1(1-\cos \psi)+k_3\sin\psi & k_2^2(1-\cos \psi)+\cos\psi & k_2k_3(1-\cos \psi)-k_1\sin\psi \\
			k_3k_1(1-\cos \psi)-k_2\sin\psi & k_3k_2(1-\cos \psi)+k_1\sin\psi & k_3^2(1-\cos \psi)+\cos\psi
		\end{pmatrix},
		\label{eqn:mrot}
	\end{align}
\end{widetext}
where $\mathbf k = (k_1,k_2,k_3)^{\text T}$ is the normalized wavevector of light. This matrix is convenient to define the polarization vectors of the ingoing and outgoing electric fields relative to the "initial" polarization vector $x$ at the polarization angle $\psi = 0$. $x = (x_1, x_2, x_3)^{\text T}$ has to be chosen according to the special crystal orientation considered, e.g. $x = (0,0,1)^{\text T}$ or $x = (1,1,0)^{\text T}/\sqrt 2$ for $\mathbf k = (1,- 1,0)^{\text T}/\sqrt 2$. For the general case we now distinguish between the ingoing polarization angle $\psi$ and the outgoing polarization angle $\varphi$ and thus get two polarization vectors $\mathbf E^{\omega}(\psi)$ and $\mathbf E^{2\omega}(\varphi)$, both of which are gained from the rotation matrix $\mathbf M_{\text{rot}}(\mathbf k, \psi)$ and the same $x$ vector by 

\begin{align}
	\mathbf E^{\omega}(\psi) &= \bigl(u(\psi),v(\psi),w(\psi)\bigr)^{\text T}& = \mathbf M_{\text{rot}}(\mathbf k, \psi) \cdot x \label{eqn:Eomega}, \\
	\mathbf E^{2\omega}(\varphi) &= \bigl(m(\varphi),n(\varphi),o(\varphi)\bigr)^{\text T}& = \mathbf M_{\text{rot}}(\mathbf k, \varphi) \cdot x .
	\label{eqn:E2omega}
\end{align}

Then we proceed as in Ref.~\cite{Mund1}. For the excitation of even-parity excitons in Fig.~\ref{fig:scheme}(a) the combination of the dipole operators for the ingoing photons is given by the symmetrized combinations of coordinate products ($e_y e_z + e_z e_y$, \ldots)
\begin{equation}
	O_{\text{DD}}(\psi)=\sqrt 2 
	\begin{pmatrix}
		v(\psi)w(\psi)\\
		u(\psi)w(\psi)\\
		u(\psi)v(\psi)
	\end{pmatrix}.
	\label{eqn:ODD}
\end{equation}
For the outgoing photon, the $\Gamma^+_5$ quadrupole operator is given by the symmetrized combinations of coordinate products ($k_y e_z + k_z e_y$, \ldots)
\begin{equation}
	O_{\text{Q5}}(\mathbf k, \varphi)=\frac{1}{\sqrt 2}
	\begin{pmatrix}
		k_2o(\varphi)+k_3n(\varphi)\\
		k_3m(\varphi)+k_1o(\varphi)\\
		k_1n(\varphi)+k_2m(\varphi)
	\end{pmatrix}
	=\begin{pmatrix}
		O_{\text{Q5,1}}(\mathbf k, \varphi)\\
		O_{\text{Q5,2}}(\mathbf k, \varphi)\\
		O_{\text{Q5,3}}(\mathbf k, \varphi)
	\end{pmatrix}.
	\label{eqn:OQ}
\end{equation} 
For the SHG intensity of the even parity excitons one thus gets 
\begin{equation}
	I^{2\omega}_{\text{even}} (\mathbf k , \psi, \varphi) \propto \left| O_{\text{DD}}(\psi) O_{\text{Q5}}(\mathbf k, \varphi) \right|^2.
	\label{eqn:Ieven}
\end{equation}
For the odd parity exciton states ($P$-excitons) in Fig.~\ref{fig:scheme}(b) the operator for the ingoing photons is given by
\begin{equation}
	O_{\text{DQ5}}(\mathbf k, \psi)=\frac{1}{\sqrt 2}
	\begin{pmatrix}
		O_{\text{Q5,3}}(\mathbf k, \psi)v(\psi)+O_{\text{Q5,2}}(\mathbf k, \psi)w(\psi)\\
		O_{\text{Q5,1}}(\mathbf k, \psi)w(\psi)+O_{\text{Q5,3}}(\mathbf k, \psi)u(\psi)\\
		O_{\text{Q5,2}}(\mathbf k, \psi)u(\psi)+O_{\text{Q5,1}}(\mathbf k, \psi)v(\psi)
	\end{pmatrix}
	\label{eqn:OTPDQ}
\end{equation}
and the operator for the outgoing photon has the representation
\begin{equation}
	O_{\text{D}}(\varphi)= 
	\begin{pmatrix}
		m(\varphi)\\
		n(\varphi)\\
		o(\varphi)
	\end{pmatrix}.
	\label{eqn:OD}
\end{equation}
For the SHG intensity of the odd parity excitons one thus gets
\begin{equation}
	I^{2\omega}_{\text{odd}} (\mathbf k,\psi, \varphi) \propto \left| O_{\text{DQ5}}(\mathbf k, \psi) O_{\text{D}}(\varphi) \right|^2.
	\label{eqn:Iodd}
\end{equation}

\begin{figure}[h]
	\begin{center}
		\includegraphics[width=0.44\textwidth]{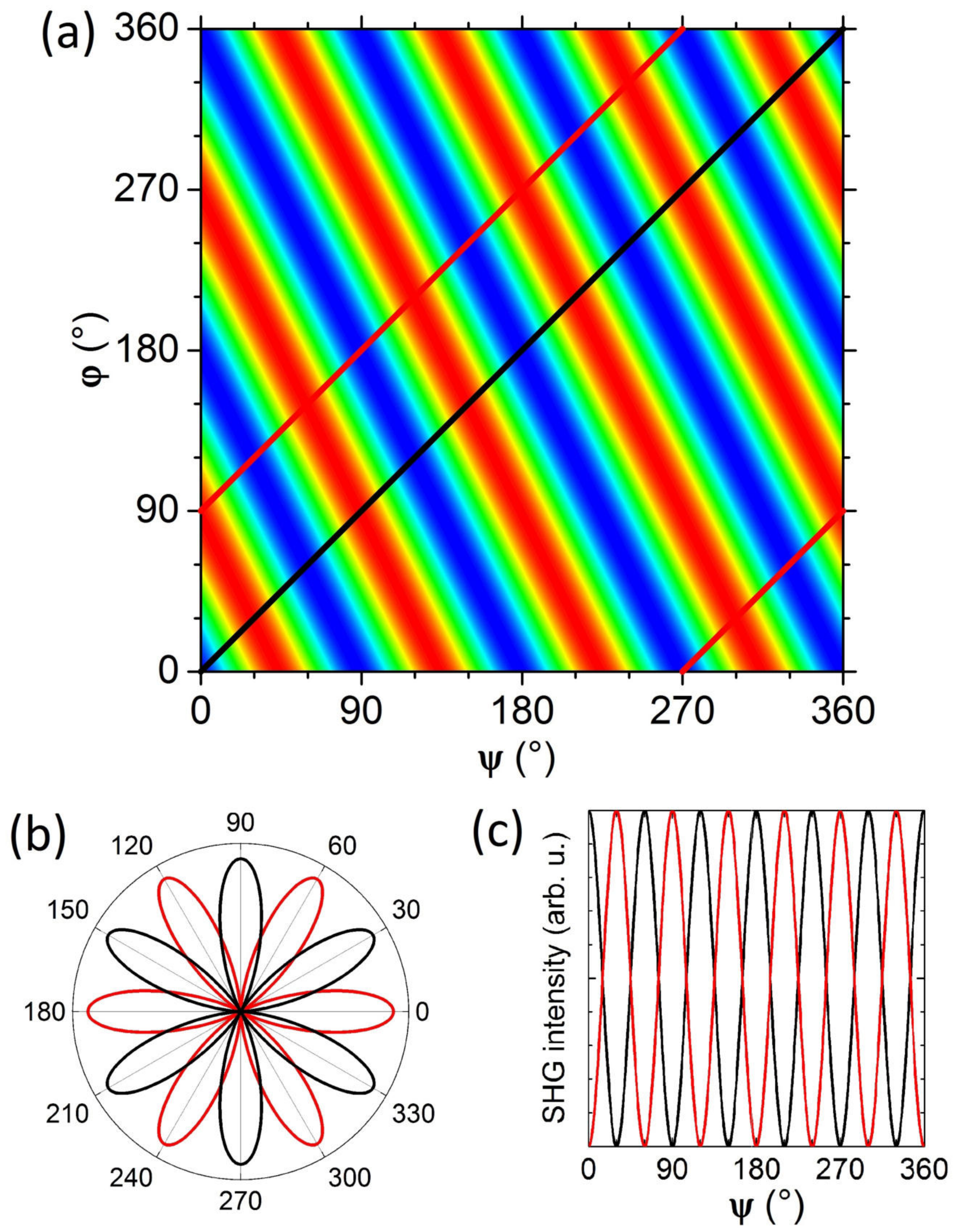} 
		\caption[Scan]{SHG intensity in dependence of the linear polarization angles of the ingoing ($\psi$) and outgoing ($\varphi$) light for the crystallographic contribution in the configuration $\mathbf k \parallel [111],x \parallel [1\bar10]$, calculated by Eq.~\eqref{eqn:Ieven}. (a) $2$D polarization dependence: Dark blue regions correspond to zero SHG intensity and red regions to maximum SHG intensity. The marked tuning lines represent the parallel $\psi=\varphi$ (black) and crossed $\psi=\varphi+90^\circ$ (red) linear polarization configurations. The SHG intensity along these tuning lines is plotted in polar representation in (b), as it was used in Refs.~\cite{Mund1, Mund2}. We prefer the Cartesian representation (c).}
		\label{fig:111}
	\end{center}
\end{figure}

Since the SHG intensities, Eqs.~\eqref{eqn:Ieven} and \eqref{eqn:Iodd}, depend on the two angles $\psi$ and $\varphi$, we plot the angular dependence of $I^{2\omega}_{\text{even, odd}} (\psi, \varphi)$ in $2$D-diagrams.
In Ref.~\cite{Mund1} the polarization dependences were applied to the SHG-allowed orientations ($\mathbf k \parallel [111]$ and $\mathbf k \parallel [11\bar 2]$) but only for the special polarization configurations $\mathbf E^\omega\parallel \mathbf E^{2\omega}$ and $\mathbf E^\omega\perp \mathbf E^{2\omega}$. In Figure~\ref{fig:111} we show the $2$D-plot for the even parity excitons [Eq.~\eqref{eqn:Ieven}] in the $\mathbf k \parallel [111]$ configuration. The odd parity excitons [Eq.~\eqref{eqn:Iodd}] show exactly the same polarization dependence, as may be expected because for both types of excitons the SHG involves in total two electric dipole and one electric quadrupole transition, albeit in different order, which, however, is not reflected by the intensity. The formerly considered selected polarization configurations are marked by the red and black, so-called tuning lines.
In addition, $1$D polar and Cartesian plots as function of the angle $\psi$ indicating the polarization of the exciting laser are given for parallel (black) and perpendicular (red) linear polarization of the fundamental and the SHG light, which resemble the tuning lines. 

Alternately, one can fix one of the polarization angles and vary the other one. When taking corresponding cuts, one still observes oscillatory behaviors but with varying period. For example, when the polarization of the ingoing fundamental light is fixed ($\psi =$ const.), the SHG intensity shows oscillations as function of the SHG polarization angle $\phi$ with a period of 180$^{\circ}$, while for fixed $\phi$ the oscillation period in $\psi$ is 
90$^{\circ}$.

To visualize, how the polarization dependence changes, when the $k$-vector is rotated continuously we present an animated contour plot (first animation, see Ref.~[42]). It starts with the polarization dependence of even excitons without external field application [Eq.~\eqref{eqn:Ieven}] for the orientation $\mathbf k \parallel [111]$ and $x \parallel [1\bar10]$ as shown in Fig.~\ref{fig:111}. In the animation the $k$-vector is rotated about the horizontal direction $x$ by $\theta=360{^\circ}$. The change of the incidence direction is accompanied by strong changes and distortions of the contour plot, from which vice versa the optical configuration can be assessed.

We now derive the SHG contributions for the ZE and MSE, which are sketched in Figs.~\ref{fig:scheme}(c) and \ref{fig:scheme}(d). The Zeeman operator transforms as $\Gamma^+_4$ (axial vector operator $\mathbf B=(B_1, B_2, B_3)^{\text T}$). We thus couple the two-photon excited $\Gamma^+_5$ exciton states ($O_{\text{DD}}$ operator of the $\Gamma_5^+$ excitons) to the ZE operator by use of Ref.~\cite{Koster} and further to the even parity operator
\begin{equation}
	O_{\text{BDD}} (\psi) =\frac{1}{\sqrt 2}
	\begin{pmatrix}
		B_2u(\psi)v(\psi)-B_3w(\psi)u(\psi)\\
		-B_1w(\psi)u(\psi)+B_3v(\psi)w(\psi)\\
		B_1u(\psi)v(\psi)-B_2v(\psi)w(\psi)
	\end{pmatrix}.
	\label{eqn:OBDD}
\end{equation}
With the quadrupole operator $O_{\text{Q5}}(\mathbf k, \varphi)$, Eq.~\eqref{eqn:OQ}, for the outgoing photon we thus get for the ZE-induced SHG signal
\begin{equation}
	I^{2\omega}_{\text{ZE}}(\mathbf k, \psi,\varphi) \propto \left| O_{\text{BDD}}(\psi) O_{\text{Q5}}(\mathbf k, \varphi) \right|^2.
	\label{eqn:IB}
\end{equation} 

Next we turn to the phenomenological description of the magneto-Stark effect. The MSE operator transforms as $\Gamma^-_4$ (polar vector operator $\mathbf E_{\text{MSE}}=(E_1, E_2, E_3)^{\text T} \propto [\bf k \times \bf B]$). We first couple the $O_{\text{DD}}(\psi)$ operator to the odd parity operator
\begin{equation}
	O_{\text{EDD}} (\psi) =\frac{1}{\sqrt 2}
	\begin{pmatrix}
		E_2u(\psi)v(\psi)+E_3w(\psi)u(\psi)\\
		E_1w(\psi)u(\psi)+E_3v(\psi)w(\psi)\\
		E_1u(\psi)v(\psi)+E_2v(\psi)w(\psi)
	\end{pmatrix}.
	\label{eqn:OEDD}
\end{equation} 
With the dipole operator $O_{\text{D}}(\varphi)$ we get for the MSE-induced SHG
\begin{equation}
	I^{2\omega}_{\text{MSE}}(\psi,\varphi) \propto \left| O_{\text{EDD}}(\psi) O_{\text{D}}(\varphi) \right|^2.
	\label{eqn:IE}
\end{equation} 

By proper choice of the polarization configuration, one can distinguish between both mechanisms. For the general case, however, one has to take into account interference effects and thus add the amplitudes in Eqs.~\eqref{eqn:IB} and \eqref{eqn:IE} before taking the squared modulus. This leads to the total intensity (where the crystallographic SHG is neglected which is always possible for rather high symmetry crystal orientations)
\begin{equation}
	\begin{gathered}
		I^{2\omega}_{\text{ZE+MSE}}(\mathbf k, \psi,\varphi) \propto \bigl| \bigl( \alpha O_{\text{BDD}}(\psi) O_{\text{Q5}}(\mathbf k,\varphi) \\
		+ \beta O_{\text{EDD}}(\psi) O_{\text{D}}(\varphi)\bigr)/\sqrt{\alpha^2+\beta^2} \bigr| ^2.
		\label{eqn:Itot}
	\end{gathered}
\end{equation} 
where $\alpha$ and $\beta$ are parameters, which have to be calculated from the appropriate interaction matrix elements, see Sec.~\ref{sec:micro}. 

\begin{figure}[h]
	\begin{center}
		\includegraphics[width=0.4\textwidth]{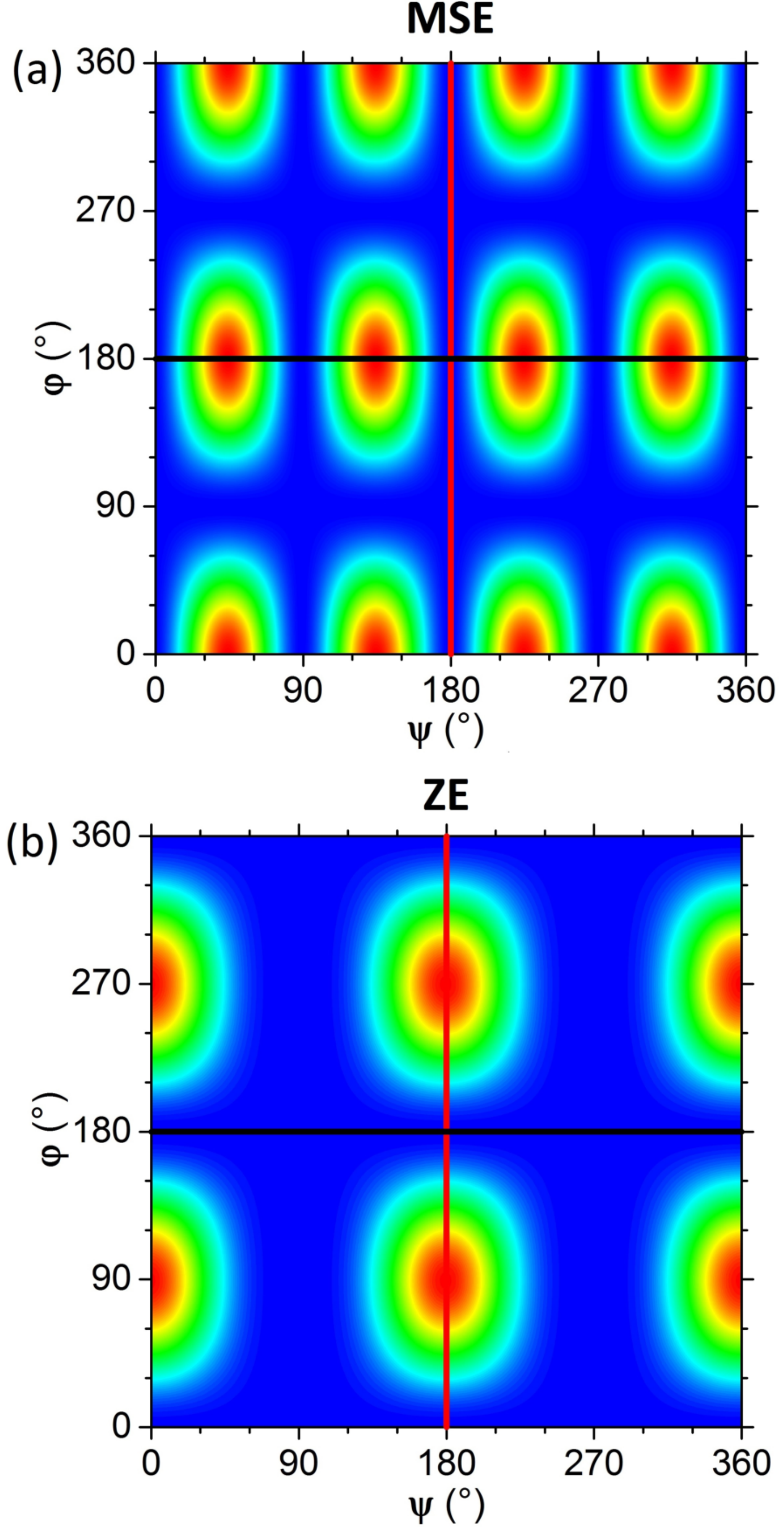}
		\caption[Scan]{$2$D polarization dependence of the SHG intensity for the configuration $\mathbf k\parallel[1\bar10],\mathbf B\parallel[110]$; (a) magneto-Stark effect [Eq.~\eqref{eqn:IE}] and (b) Zeeman effect [Eq.~\eqref{eqn:IB}].}		
		\label{fig:gamma5}
	\end{center}
\end{figure}

For the SHG forbidden orientation $\mathbf k \parallel [1\bar10]$ we have chosen the magnetic field in Voigt configuration $\mathbf B \parallel [110]$. 
The corresponding electric field of the magneto-Stark effect is $\mathbf E_{\text{MSE}} \sim \mathbf k \times \mathbf B\parallel [001]$. In Figure~\ref{fig:gamma5} we show for this configuration the $2$D plots for the isolated ZE [Eq.~\eqref{eqn:IB}] and the isolated MSE [Eq.~\eqref{eqn:IE}]. We refer to these $2$D plots for the simulation of our experimental results in Sec. \ref{sec:experimental results}, as they help us to identify the configurations in which only one effect contributes. As shown in Fig.~\ref{fig:gamma5}(a) and \ref{fig:gamma5}(b), the black $\psi$-tuning line for $\varphi = 180^\circ$ exhibits maximum SHG with a $90^\circ$ period for the MSE but no SHG for the ZE, whereas the red $\varphi$-tuning line for $\psi = 180^\circ$ exhibits maximum SHG with a $180^\circ$ period for the ZE but no SHG for the MSE. The corresponding experimental results are shown in Sec.~\ref{sec:experimental results} Figs.~\ref{fig:MSE_experiment} and \ref{fig:ZE_experiment}.

Let us now address the selection rules for two-photon absorption (TPA).
From the detailed SHG polarization dependences for the different cases [Eqs.~\eqref{eqn:Ieven}, \eqref{eqn:Iodd}, \eqref{eqn:IB}, \eqref{eqn:IE} and \eqref{eqn:Itot}] one can easily derive the equivalent polarization dependences for TPA, which only depend on the polarization angle $\psi$ of the ingoing photons, by merely omitting in the equations the outgoing operator describing either a quadruploe or a dipole transition [$O_{\text{Q5}}(\mathbf k, \varphi)$, $O_{\text{D}}(\varphi)$]. Experimentally, TPA is monitored by photoluminescence excitation spectroscopy detecting the emission of a photon from a state into the electron-hole pair has relaxed after excitation by the two photon transition (e.g., in our case the spectrometer is set to detection at the energy of the $1S$ exciton or its $\Gamma_3^-$ phonon replica). During relaxation the coherence excited in the system by optical excitation is typically destroyed.

We thus get for the TPA polarization dependence of the $\Gamma_5^+$ excitons from Eq.~\eqref{eqn:ODD} 
\begin{equation}
	I^{2\omega}_{\text{TPA}}(\psi) \propto \left| O_{\text{DD}}(\psi)\right|^2.
	\label{eqn:ITPA}
\end{equation} 
The resulting polarization dependence of TPA will be discussed below in combination with corresponding experimental data, shown in Sec.~\ref{sec:experimental results}. Importantly, TPA is allowed along the direction $\mathbf k \parallel [1\bar10]$, where SHG is forbidden at zero magnetic field as the coherent photon emission is blocked.

\subsection{Weaker Processes}
\label{sec:weak}

In this section we consider the weaker SHG processes as addressed at the end of Sec.~\ref{subsec:bands}. Namely, we address the excitons where the two-photon excitation channel is significantly suppressed as compared to the $\Gamma_5^+$ $S$ and $D$ excitons mixed by the exchange interaction. These are the odd-parity $P$-excitons ($\Gamma_4^-$ representation) whose two-photon excitation requires a quadrupolar process and the $D$ excitons of $\Gamma_1^+$ and $\Gamma_3^+$ symmetries which are decoupled from the $S$-excitons. 

\begin{figure}[h]
	\begin{center}
		\includegraphics[width=0.48\textwidth]{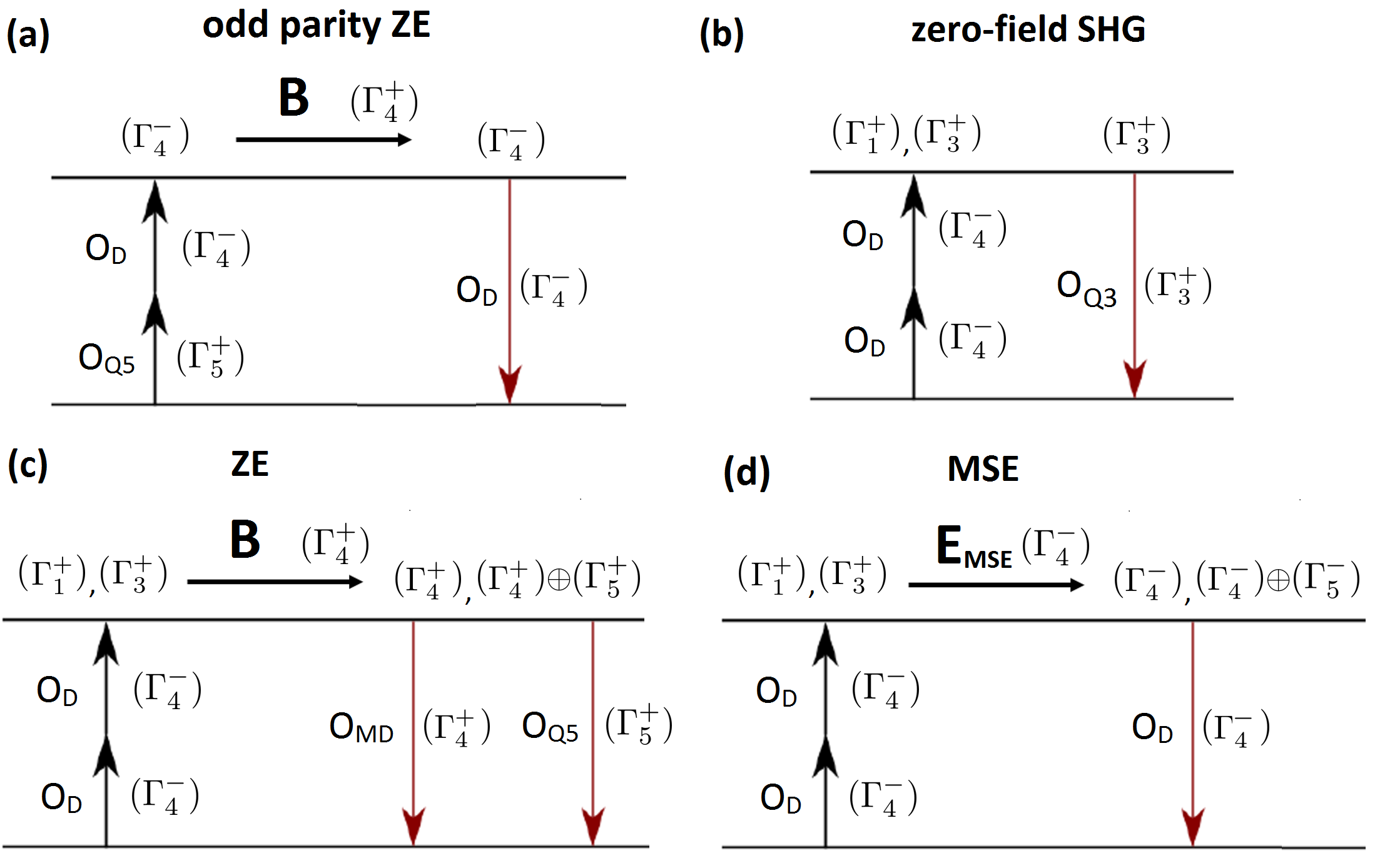}
		\caption[Scan]{Schematics of the weaker SHG processes: (a) ZE of odd parity excitons, (b) zero-field two-photon excitation of the $\Gamma_1^+$ and $\Gamma_3^+$ $D$-exciton states, (c) ZE- and (d) MSE-induced SHG transitions, respectively, on the $\Gamma_1^+$ and $\Gamma_3^+$ $D$-excitons.}
		\label{fig:scheme_gamma31}
	\end{center}
\end{figure}

\begin{figure}[h]
	\begin{center}
		\includegraphics[width=0.375\textwidth]{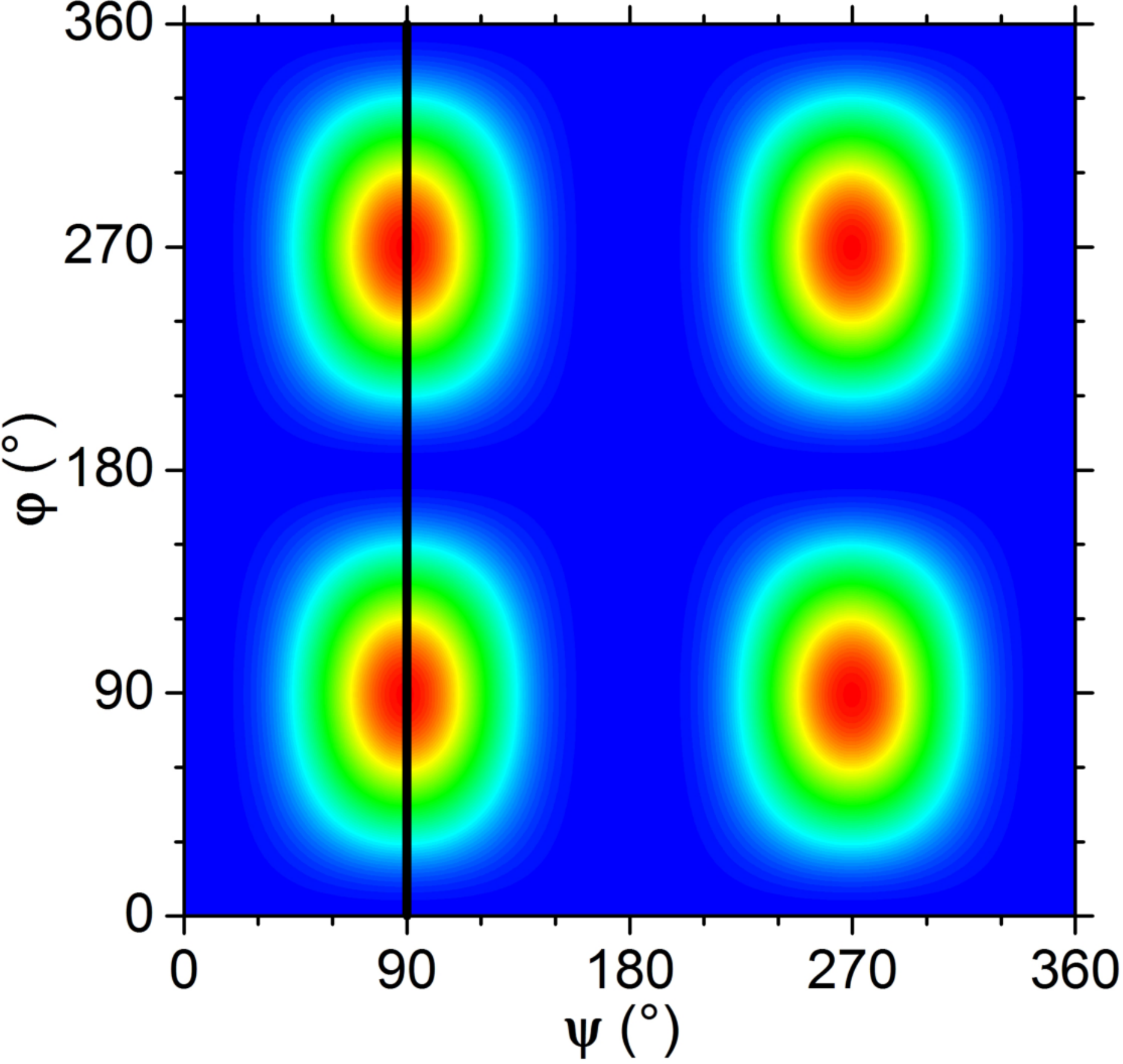}
		\caption[Scan]{$2$D plot of the ZE-SHG from the odd-parity $P$-excitons [Eq.~\eqref{eqn:IBDQ5}, Fig.~\ref{fig:scheme_gamma31}(a)] for the forbidden configuration $\mathbf k \parallel [1\bar10]$, $\mathbf B \parallel [110]$, the black line corresponds to tuning of the outgoing polarization $\varphi$ for fixed ingoing polarization $\psi = 90^\circ$.}		
		\label{fig:gamma4sim}
	\end{center}
\end{figure}

The various scenarios for the weaker processes are sketched in Fig.~\ref{fig:scheme_gamma31}, for the $P$-excitons in (a) at non-zero magnetic field, and for the $D$ excitons in (b) in zero field as well as in (c,d) for a finite field, activating the ZE and the MSE. We start with Eq.~\eqref{eqn:OTPDQ} for the derivation of the ZE of the odd-parity $P$ excitons. The operator $O_{\text{DQ5}}(\mathbf k,\psi)$ is coupled by the Zeeman operator ($\Gamma_4^+$) to a $\Gamma_4^-$ operator for the outgoing dipole transition, which leads to
\begin{align}
	O_{\text{BDQ5}} (\psi) = \frac{1}{\sqrt 2}
	\begin{pmatrix}
		B_2O_{\text{DQ5,3}}(\mathbf k, \psi)-B_3O_{\text{DQ5,2}}(\mathbf k,\psi)\\
		-B_1O_{\text{DQ5,3}}(\mathbf k,\psi)+B_3O_{\text{DQ5,1}}(\mathbf k,\psi)\\
		B_1O_{\text{DQ5,2}}(\mathbf k,\psi)-B_2O_{\text{DQ5,1}}(\mathbf k,\psi)
	\end{pmatrix}.
	\label{eqn:OBDQ5}
\end{align}
The $O_{\text{DQ5,i}}$ ($i=1,2,3$) are the components of the vector in Eq.~\eqref{eqn:OTPDQ}.
With the dipole operator $O_{\text{D}}(\varphi)$, Eq.~\eqref{eqn:OD}, one gets
\begin{equation}
	I^{2\omega}_{\text{BDQ}}(\mathbf k,\psi,\varphi) \propto \left| O_{\text{BDQ5}}(\mathbf k,\psi)\cdot O_{\text{D}}(\varphi) \right|^2
	\label{eqn:IBDQ5}
\end{equation} 
for the ZE-induced SHG of the odd parity excitons.
In Figure~\ref{fig:gamma4sim} we show the $2$D plot for the ZE of the odd parity excitons.
By proper choice of the tuning line one can selectively excite the ZE of the $P$ excitons and thus suppress the potentially dominant excitation of the ZE and the MSE of even parity excitons [Fig.~\ref{fig:gamma5}]. E.g., one can set the ingoing linear polarization to $\psi = 90^{\circ}$ and vary the detection angle $\varphi$ of the second harmonic light. For this configuration both the ZE- and the MSE-induced SHG of the $\Gamma_5^+$ states disappear. 

We now turn to the analysis of the SHG effect on the even parity $\Gamma_1^+$ and $\Gamma_3^+$ $D$-excitons in zero field as well as in a magnetic field. In the zero-field case only a SHG signal from the $\Gamma_3^+$ excitons is expected by quadrupole emission. In magnetic field, however, we expect for the ZE besides the electric quadrupole ($\Gamma_5^+$) also a magnetic dipole contribution of $\Gamma_4^+$ symmetry. For the MSE we have to consider only the odd-parity dipole operator of $\Gamma_4^-$ symmetry. The relevant excitation and emission operators are derived according to Ref.~\cite{Koster}. For the $\Gamma_1^+$ and $\Gamma_3^+$ excitation operators $O_{\text{DD1}}$ and $O_{\text{DD3}}$ we thus get:
\begin{align}
	O_{\text{DD1}}(\psi)&=\frac{1}{\sqrt 3},
	\\
	O_{\text{DD3}}(\psi)&=\frac{1}{\sqrt 6} 
	\begin{pmatrix}
		-u(\psi)^2-v(\psi)^2+2w(\psi)^2\\
		-\sqrt 3 u(\psi)^2 -\sqrt 3 v(\psi)^2
	\end{pmatrix}
	\\&=
	\begin{pmatrix}
		O_{\text{DD3,1}}\\
		O_{\text{DD3,2}}
	\end{pmatrix}.\notag
	\label{eqn:ODD}
\end{align}

For the outgoing photons we now consider besides the $\Gamma_5^+$ quadrupole operator, which is treated in the previous section in Eq.~\eqref{eqn:OQ}, also the $\Gamma_4^+$ magnetic dipole operator and the $\Gamma_3^+$ quadrupole operator
\begin{equation}
	O_{\text{MD}}(\mathbf k, \varphi)=\frac{1}{\sqrt 2} 
	\begin{pmatrix}
		k_2 o(\varphi)-k_3 n(\varphi)\\
		- k_1 o(\varphi) + k_3 m(\varphi)\\
		k_1 n(\varphi) - k_2 m(\varphi)
	\end{pmatrix},
	\label{eqn:OMD3}
\end{equation}
\begin{equation}
O_{\text{Q3}}(\mathbf k, \varphi)=\frac{1}{\sqrt 6} 
\begin{pmatrix}
-k_1m(\varphi)-k_2n(\varphi)+2 k_3 o(\varphi)\\
-\sqrt 3 k_1 m(\varphi) -\sqrt 3 k_2 n(\varphi)
\end{pmatrix}.
\label{eqn:OQ3}
\end{equation}
These operators are easily derived from Ref.~\cite{Koster} by considering the direct product of the $\mathbf k$-vector and the polarization vector, both of which are of $\Gamma_4^-$ symmetry:
\begin{equation}
	\Gamma_4^-\otimes\Gamma_4^-=\Gamma_1^+\oplus\Gamma_3^+\oplus\Gamma_4^+\oplus\Gamma_5^+,
\end{equation}
where the $\Gamma_1^+$ contribution vanishes because the $k$-vector and polarization vector are orthogonal to each other. 
For the SHG intensity in zero field only the quadrupole operator leads to a signal, because there is no two-photon excitable $\Gamma^+_4$ state for two identical exciting photons:
\begin{align}
	I^{2\omega}_{\text{even Q3}} (\mathbf k , \psi, \varphi) \propto \left| O_{\text{DD3}}(\psi) \cdot O_{\text{Q3}}(\mathbf k, \varphi) \right|^2.
\end{align}

We now proceed with the ZE and the MSE, as in the previous section for the dominant processes. The relevant processes are sketched in Fig.~\ref{fig:scheme_gamma31}(c) and \ref{fig:scheme_gamma31}(d). We thus couple the two-photon excited $\Gamma^+_1$ and $\Gamma^+_3$ exciton states to the ZE operator $\Gamma^+_4$ and further to the even parity operator:
\begin{align}
	O_{\text{BDD1}} (\psi) &=\frac{1}{\sqrt 3}
	\begin{pmatrix}
		B_1\\
		B_2\\
		B_3
	\end{pmatrix}.
	\\
	O_{\text{BDD3to5}} (\psi) &=\frac{1}{2\sqrt 6}
	\begin{pmatrix}
		-\sqrt 3B_1 O_{\text{DD3,1}}-B_1O_{\text{DD3,2}}\\
		\sqrt 3B_2O_{\text{DD3,1}}-B_2O_{\text{DD3,2}}\\
		2B_3	O_{\text{DD3,2}}
	\end{pmatrix}.
	\label{eqn:OEDD3to5}
	\\
	O_{\text{BDD3to4}} (\psi) &=\frac{1}{2\sqrt 6}
	\begin{pmatrix}
		- B_1O_{\text{DD3,1}}+\sqrt 3B_1O_{\text{DD3,2}}\\
		-B_2O_{\text{DD3,1}}-\sqrt 3B_2O_{\text{DD3,2}}\\
		2B_3	O_{\text{DD3,1}}
	\end{pmatrix}.
	\label{eqn:OEDD3to4}
\end{align}
With the magnetic dipole operator $O_{\text{MD}}(\mathbf k, \varphi)$ [Eq.~\eqref{eqn:OMD3}] and the electric quadrupole operator $O_{\text{Q5}}(\mathbf k, \varphi)$ [Eq.~\eqref{eqn:OQ}] for the outgoing photon we obtain for the ZE-induced SHG:
\begin{align}
	I^{2\omega}_{\text{B1}}(\mathbf k, \psi,\varphi) &\propto \left| O_{\text{BDD1}}(\psi) O_{\text{MD}}(\mathbf k, \varphi) \right|^2,
	\label{eqn:IB1}
	\\
	I^{2\omega}_{\text{B3to5}}(\mathbf k, \psi,\varphi) &\propto \left| O_{\text{BDD3to5}}(\psi) O_{\text{Q5}}(\mathbf k, \varphi) \right|^2,
	\label{eqn:IB35}
	\\
	I^{2\omega}_{\text{B3to4}}(\mathbf k, \psi,\varphi) &\propto \left| O_{\text{BDD3to4}}(\psi) O_{\text{MD}}(\mathbf k, \varphi) \right|^2.
	\label{eqn:IB34}
\end{align} 
We couple the $O_{\text{DD}}(\psi)$ operator to the odd parity $\Gamma_4^-$ MSE operator and get:
\begin{align}
	O_{\text{EDD1}} (\psi) &=\frac{1}{\sqrt 3} 
	\begin{pmatrix}
		E_1\\
		E_2\\
		E_3
	\end{pmatrix}.
	\label{eqn:OEDD1}\\
	O_{\text{EDD3}} (\psi) &=\frac{1}{2\sqrt 6}
	\begin{pmatrix}
		- E_1O_{\text{DD3,1}}+\sqrt 3E_1O_{\text{DD3,2}}\\
		-E_2O_{\text{DD3,1}}-\sqrt 3E_2O_{\text{DD3,2}}\\
		2E_3	O_{\text{DD3,1}}
	\end{pmatrix}.
	\label{eqn:OEDD3}
\end{align} 
With the dipole operator $O_{\text{D}}(\varphi)$ we obtain for the MSE-induced SHG intensity:
\begin{align}
	I^{2\omega}_{\text{E1}}(\psi,\varphi) &\propto \left| O_{\text{EDD1}}(\psi) O_{\text{D}}(\varphi) \right|^2,
	\label{eqn:IE1}\\
	I^{2\omega}_{\text{E3}}(\psi,\varphi) &\propto \left| O_{\text{EDD3}}(\psi) O_{\text{D}}(\varphi) \right|^2.
	\label{eqn:IE3}
\end{align} 

\begin{figure}[h]
	\begin{center}
		\includegraphics[width=0.4\textwidth]{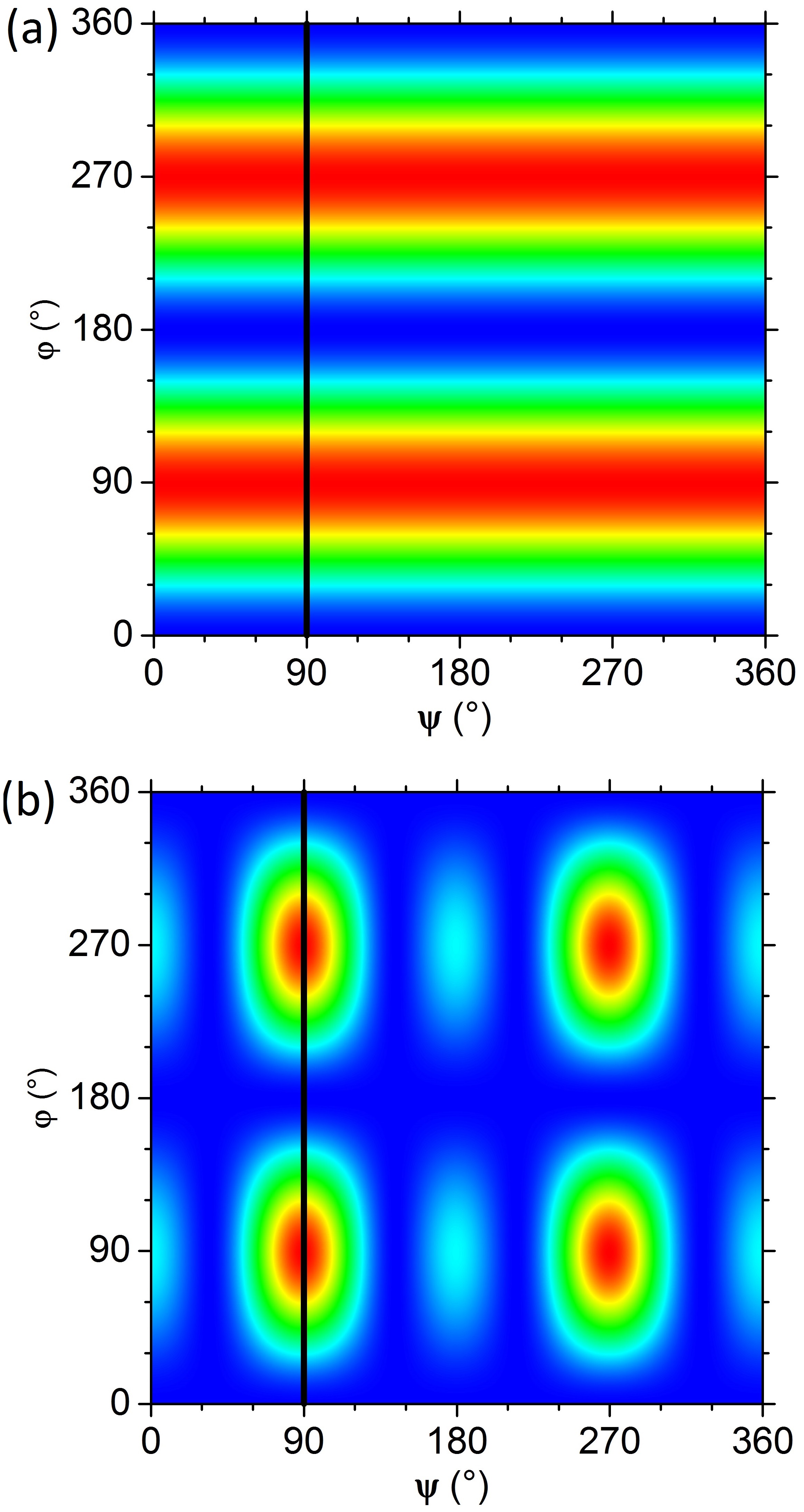}
		\caption[Scan]{$2$D plots of weaker SHG processes for the forbidden configurations $\mathbf k \parallel [1\bar10]$, $\mathbf B \parallel [110]$: (a) ZE- and MSE-induced SHG of $\Gamma_1^+$ excitons [Eqs.~\eqref{eqn:IB1} and \eqref{eqn:IE1}, Fig.~\ref{fig:scheme_gamma31}(c) and \ref{fig:scheme_gamma31}(d)], (b) ZE- and MSE-induced SHG of $\Gamma_3^+$ excitons [Eq.~\eqref{eqn:IB35}, \eqref{eqn:IB34} and \eqref{eqn:IE3}, Fig.~\ref{fig:scheme_gamma31}(c) and \ref{fig:scheme_gamma31}(d)].}	
		\label{fig:gamma31sim}
	\end{center}
\end{figure}

In Figure~\ref{fig:gamma31sim} we show the associated $2$D plots [Eqs.~\eqref{eqn:IB1} to \eqref{eqn:IB34} and Eqs.~\eqref{eqn:IE1} and \eqref{eqn:IE3}], again for the configuration $\mathbf k \parallel [1\bar1 0]$, $\mathbf B \parallel [110]$ and thus $\mathbf E_{\text{MSE}} \parallel [001]$. Compared to the preceding plots, some interesting features are seen: For varying the detection angle $\varphi$ at constant $\psi$-polarization one observes an oscillatory behavior of the intensity with period of 180$^{\circ}$. However, when varying $\psi$ pronounced significant differences show up. Namely, for a fixed $\varphi$ the SHG induced by the ZE does not depend at all on the linear polarization of the fundamental light. On the other hand, for the MSE-induced SHG one observes in that case pronounced changes which do not correspond to a simple harmonic oscillation, but the amplitude is strongly modulated leading to a periodicity in $\psi$ of 180$^{\circ}$ and not of 90$^{\circ}$. 
Note that also here unique configurations can be found which allow not only distinction of the ZE and the MSE for the $\Gamma_1^+$ and $\Gamma_3^+$ excitons, but also distinction from the processes related to the $\Gamma_5^+$ excitons. This is obvious for the ZE with its insensitivity to $\psi$, but also for the MSE with the appearance of SHG for $\psi$ = 90$^{\circ}$ and 180$^{\circ}$ with strongly different strengths.
As in the previous section one can derive the equivalent polarization dependences for TPA by merely omitting in the SHG equations the outgoing operator [$O_{\text{MD}}(\mathbf k, \varphi)$, $O_{\text{Q3}}(\varphi)$, $O_{\text{D}}(\varphi)$].

In Appendix \ref{appendix:polarization} we present $2$D polarization diagrams for eight selected crystalline orientations in zero field and for two magnetic field orientations (Voigt and Faraday configuration).

\FloatBarrier

\section{Microscopic theory}
\label{sec:micro}

In the presence of an external electromagnetic field the electron momentum operator $\widehat{\mathbf p} =- \mathrm i \hbar \mathbf \nabla$ is replaced by $\widehat{\mathbf p}-e\mathbf A/c$, where $\mathbf A$ is the vector potential of the field. Hereafter we use the gauge, where the scalar potential of the light wave is zero. Thus, the light-matter interaction operator assumes the form
\begin{equation}\label{operator}
	\widehat{V}=-\frac{e}{c m_0}{\widehat{\mathbf p}\cdot\mathbf A},
\end{equation}
where $m_0$ is the free electron mass; note that the quadratic in $\mathbf A$ term plays no role for interband transitions. For plane monochromatic waves, the complex amplitudes of the vector potential and the electric field $\mathbf A, \mathbf E \propto \exp{(\mathrm i \mathbf q\mathbf r - \mathrm i \omega t)}$ are interrelated by $\mathbf E = \mathrm i \omega\mathbf A /c$. Also, the induced dielectric polarization and electric current density at double fundamental frequency are related as $\mathbf j = -2\mathrm i \omega \mathbf P$, which makes it possible to recast the second harmonic susceptibility $\chi_{ikl}(\mathbf k, \mathbf B)$ in the general phenomenological relation [cf. Eqs.~\eqref{SHG:cryst} and \eqref{SHG:B}]
\[
\mathbf
P_i = \chi_{ikl}(\mathbf k, \mathbf B) E_k E_l
\]
as [cf.~\cite{Glazov2017}]
\begin{equation}\label{chi_gen}
	\chi_{ikl}(\mathbf k, \mathbf B) = \Xi \sum_{x,s}\frac{\langle 0| \widehat{p}_i |x\rangle \langle x| \widehat{p}_k|s\rangle \langle s| \widehat{p}_l|0\rangle}{(2\hbar\omega-E_x)(\hbar\omega-E_s)}.
\end{equation}
Here we introduce explicitly the dependence of the susceptibility on the wavevector of light and the static external magnetic field, $\Xi = {e^3}/{(2\mathrm i m_0^3\omega^3)}$, $s$ enumerates the intermediate states of the crystal, $E_s$ is the energy of the state $s$, and $x$ enumerates the exciton (final) states for the two-photon absorption, $E_x$ is the energy of the exciton state. It is noteworthy that the Coulomb interaction between the electron and hole in the intermediate states can be disregarded provided that the exciton binding energy is much smaller than $\hbar\omega$. Equation~\eqref{chi_gen} clearly shows that in a centrosymmetric crystal at $\mathbf k=0$ SHG is forbidden because the states $x$ and $s$ have definite parity, while to contribute to Eq.~\eqref{chi_gen}, the given excitonic state $x$ should be simultaneously active in two photon absorption (i.e., be even at space inversion) and in one photon emission (i.e., be odd at space inversion). This is possible only if the wavevector of radiation is taken into account. In what follows we take into account only $\mathbf k$-linear contributions in Eq.~\eqref{chi_gen}. \\

\subsection{SHG in the absence of a magnetic field}
\label{sec:micro:B0}

At $B=0$, there is only one independent contribution to the susceptibility, Eq.~\eqref{SHG:cryst:1}. It vanishes if light is propagating along one of the cubic axes and also if the light is propagating along one of the $\langle 110\rangle$ axes~\cite{Mund2}. SHG is most prominent for $\mathbf q \parallel [111]$. In the set of axes with $x'\parallel [11\bar 2]$, $y' \parallel [\bar 110]$ and $z'\parallel [111]$ we can rewrite Eq.~\eqref{SHG:cryst:1} as
\begin{subequations}
	\label{SHG:cryst:111}
	\begin{align}
		&P_{x'} = -\frac{\chi_c}{\sqrt{2}} q_{z'} (E_{x'}^2-E_{y'}^2),\\
		&P_{y'} = \frac{2\chi_c}{\sqrt{2}}q_{z'} E_{x'} E_{y'}.
	\end{align}
\end{subequations}

Let us derive the contribution of the $\Gamma_5^+$ $S$-shell excitons to $\chi_c$. These states are two-photon active. Their wavefunction can be written as
\begin{equation}
	\label{Psi:ns:5}
	\psi_{nS;\alpha\beta}(\mathbf r_e, \mathbf r_h) = e^{\mathrm i \mathbf q \mathbf r_{cm}} \Phi_{nS}(r)\mathcal U_{\alpha\beta}(\mathbf r_e, \mathbf r_h),
\end{equation}
where $\mathbf r_e$ and $\mathbf r_h$ are the electron and hole position vectors, $\mathbf r = \mathbf r_e - \mathbf r_h$ is the relative motion coordinate, $\mathbf r_{\text{cm}} = (m_e \mathbf r_e+m_h \mathbf r_h)/M$ is the center of mass coordinate with $m_e$, $m_h$ and $M=m_e+m_h$ being the electron, hole and exciton translational masses, respectively, $\Phi_{nS}(r)$ is the hydrogenic envelope with $n=1,2,\ldots$ being the principal quantum number, and $\mathcal U_{\alpha\beta}$ is the two-particle Bloch function. Here we enumerate the basic functions of the $\Gamma_5^+$ representation by the subscript $\alpha\beta$ ($\alpha \ne \beta$) running through $yz$, $zx$ and $xy$. The part of the susceptibility responsible for the two-photon absorption takes the form
\begin{multline}
	\label{M:TPA:ns:5}
	M^{(2)}_{nS;\alpha\beta;kl} = \sum_s \frac{\langle x| \widehat{p}_k|s\rangle \langle s| \widehat{p}_l|0\rangle}{\hbar\omega-E_s} \\
	= \Phi_{nS}^*(0) R [\delta_{\alpha,k}\delta_{\beta, l} + \delta_{\alpha,l} \delta_{\beta,k}],
\end{multline}
where ${R}\equiv R(\omega)$ is the parameter which includes the sum over the intermediate states (particularly, the electron states of the odd parity $\Gamma_8^-$ band) of the products of the momentum operators and the energy denominators. Similarly, quadrupole emission for the $\Gamma_5^+$ states is described by the matrix element
\begin{multline}
	\label{M:Q:ns:5}
	M^{(1)}_{nS;\alpha\beta;ij} = \langle 0| \widehat{p}_i |x\rangle \\
	= k_j \Phi_{nS}(0) Q [\delta_{\alpha,i}\delta_{\beta, j} + \delta_{\alpha,j} \delta_{\beta,i}].
\end{multline}
Here $Q$ is another parameter which accounts for the $\mathbf k\cdot \mathbf p$ mixing with the $\Gamma_8^-$ bands. As a result,
\begin{equation}
	\label{chi:c:micro}
	\chi_c \propto \Xi Q R \sum_{n} \frac{|\Phi_{nS}(0)|^2}{2\hbar\omega - E_g -E_{nS}+\mathrm i \Gamma_{nS}},
\end{equation}
where $E_{nS}<0$ is the energy of the $nS$-shell bound exciton reckoned from the electron-hole continuum and $E_g$ is the band gap. Further, we introduced a phenomenological damping $\Gamma_{nS}$ in Eq.~\eqref{chi:c:micro}. To shorten the notations a numerical factor is omitted in Eq.~\eqref{chi:c:micro}, see Appendix~\ref{appendix:toy} for details.

As a next step we evaluate SHG of the $P$-shell excitons of $\Gamma_4^-$ symmetry. In contrast to $S$-shell states, the $P$-excitons are dipole active, but require a quadrupolar transition for two-photon excitation. The calculation shows that the contribution of the $P$-shell excitons to the susceptibility can be recast as
\begin{equation}
	\label{chi:c:micro:P}
	\chi_c \propto \Xi Q R \sum_{n} \frac{|a_0\Phi'_{nP}(0)|^2}{2\hbar\omega - E_g -E_{nP}+\mathrm i \Gamma_{nP}},
\end{equation}
where $a_0$ is the effective length being on the order of the lattice constant, see Appendix~\ref{appendix:toy} and Ref.~\cite{Glazov2017} for details, $\Phi_{nP}'(0)$ is the derivative of the $P$-shell radial envelope at coinciding electron and hole coordinates, and $\Gamma_{nP}$ is the corresponding damping. Let us now compare the peak values of the second-order susceptibility at the $S$- and $P$-excitons in Cu$_2$O. It follows from Eqs.~\eqref{chi:c:micro} and \eqref{chi:c:micro:P} that this ratio is given by 
\begin{equation}
	\label{rat:SP}
	\left|\frac{\chi_c(\omega_{nS})}{\chi_c(\omega_{nP})} \right| \sim \left|\frac{\Phi_{nS}(0)}{a_0 \Phi_{nP}'(0)} \right| \frac{\Gamma_{nP}}{\Gamma_{nS}}.
\end{equation}
Making use of the explicit form of the hydrogenic wavefunctions and assuming similar dampings $\Gamma_{nS} \sim \Gamma_{nP}$ we arrive at
\begin{equation}
	\label{rat:SP:1}
	\left|\frac{\chi_c(\omega_{nS})}{\chi_c(\omega_{nP})} \right| \sim \left(\frac{a_B}{a_0}\right)^2 \frac{6n^2}{n^2-1} \sim \left(\frac{a_B}{a_0}\right)^2.
\end{equation} 
Thus, compared to the contribution of the $S$-excitons the $P$-shell states at zero magnetic field provide a parametrically smaller contribution to SHG, $\sim (a_0/a_B)^2$ where $a_B$ is the exciton Bohr radius [cf. Ref.~\cite{Glazov2017}].

Let us now turn to SHG of the $D$-shell excitons. As already discussed in Sec.~\ref{subsec:bands}, the $D$-shell excitonic states transform according to the $\Gamma_1^+$, $\Gamma_3^+$, $\Gamma_4^+$ and $\Gamma_5^+$ irreducible representations of the $O_h$ point symmetry group. The states of $\Gamma_5^+$ symmetry are efficiently mixed with the $S$-shell orthoexcitons forming a series of $S/D$-shell states~\cite{Uihlein}. Their contribution to the SHG susceptibility has the form

\begin{equation}
	\label{chi:c:micro:D:5}
	\chi_c \propto \Xi Q R \sum_{n} \frac{|a_n^2\Phi_{nD}''(0)|^2}{2\hbar\omega - E_g -E_{nD}+\mathrm i \Gamma_{nD}},
\end{equation}
where $E_{nD}$ is the energy of the $\Gamma_5^+$ $D$-shell state, $\Phi_{nD}''(0)$ is the second derivative of the $S$-exciton radial envelope at coinciding coordinates of electron and hole, $a_n$ is the mixing parameter. Strictly speaking, in Eqs.~\eqref{chi:c:micro} and \eqref{chi:c:micro:D:5} the energies of the $S/D$ mixed states should be used. Similarly, the redistribution of the oscillator strength from the $S$- to the $D$-shell excitons should be taken into account in Eq.~\eqref{chi:c:micro}. This can be done in the approach of Ref.~\cite{RommelSHG}. The mixing parameter has the dimension of a length but unlike $a_0$ in Eq.~\eqref{chi:c:micro:P} it is given by the combination of the Luttinger parameters and the spin-orbit splitting constant, as this mixing comes from the coupling between the close in energy $\Gamma_7^+$ and $\Gamma_8^+$ valence bands, see Fig.~\ref{fig:bands} and Ref.~\cite{Uihlein}. Rough estimates show that $a_n \sim a_B$, i.e., it is on the order of the exciton Bohr radius. Thus the $\Gamma_5^+$ symmetry $S$ and $D$ states provide comparable contributions to SHG.

Out of the remaining $D$-shell excitons, only those with $\Gamma_1^+$ and $\Gamma_3^+$ symmetry are active in two-photon absorption. However, unlike the $\Gamma_5^+$ states, for both the $\Gamma_1^+$ (naturally, the $\Gamma_1^+$ exciton does not contribute to $\chi_c$ in Eq.~\eqref{SHG:cryst:1} since it generates a polarization along $\mathbf k$) and the $\Gamma_3^+$ excitons, the coupling with the $S$-shell states is absent. Thus, the two-photon excitation of these $D$-shell excitons requires for transitions via intermediate states an additional $\mathbf k\cdot \mathbf p$ mixing with remote bands. For example, the $\Gamma_1^+$ exciton can be excited with two photons, the $\Gamma_1^+$ $S$-shell exciton with the hole in the $\Gamma_7^+$ valence band and with the electron in the remote $\Gamma_{7,c}^+$ conduction band, taking into account the second-order $\mathbf k\cdot\mathbf p$ mixing of the remote $\Gamma_{7,c}^+$ and the bottom $\Gamma_6^+$ conduction bands. Similarly, the $\Gamma_3^+$ states can be activated by taking into account intermediate states in the $\Gamma_{8,c}^+$ symmetry bands and the corresponding second-order $\mathbf k\cdot\mathbf p$-mixing with the $\Gamma_6^+$ conduction band. The energy separation to these bands $E_{g,remote} \sim 10$~eV~\cite{PhysRevB.21.1549}. As a result, the susceptibility acquires the form of Eq.~\eqref{chi:c:micro:D:5} but with replacing $a_n$ by a quantity $\sim a_0 \ll a_n$. This results in a significant suppression of the SHG of the $\Gamma_3^+$ $D$-shell excitons as compared to the contribution of the $\Gamma_5^+$ $S/D$-excitons.

\subsection{SHG in presence of magnetic field}
\label{sec:micro:B}

Although the magnetic field does not break the $\mathcal P$-symmetry, it is expected to produce a significant effect on SHG, see the phenomenological Eqs.~\eqref{SHG:MS} and \eqref{SHG:Z} and the discussion in Secs.~\ref{sec:dom} and \ref{sec:weak}. In the $B$-linear regime two key effects occur: (i) the Zeeman effect resulting in a splitting/mixing of different states of the same parity, e.g., mixing of a state which is (in a given field and polarization configuration) quadrupolar forbidden but active in two-photon absorption with a state which is quadrupolar active but forbidden in two-photon absorption, and (ii) the magneto-Stark effect which is a result of the combined action of the magnetic field and exciton propagation and leads to a mixing of excitons of different parity via the equivalent electric field given by $\mathbf E_{\text{MSE}} \propto [\mathbf k\times \mathbf B]$. We will illustrate these particular microscopic mechanisms considering the experimentally relevant geometry with the light propagating along the $z_1 \parallel [1\bar 10]$ axis and the magnetic field applied along the $x_1\parallel [110]$ axis, with $y_1\parallel [001]$, see Fig.~\ref{fig:orientation}. As discussed above this is the so-called forbidden geometry along which the crystalline SHG (at $B=0$) is not allowed.

\subsubsection{$\Gamma_5^+$ excitons}\label{G5:subsec}

We start the analysis with the simplest case of the $\Gamma_5^+$ excitonic states. In the studied geometry the triplet of the $\Gamma_5^+$ $S/D$-mixed states can be described by the wavefunctions $\tilde \Psi_{1,2,3}$ which transform as
\begin{equation}
	\label{5:110}
	\tilde \Psi_1 \propto \frac{x_1^2-z_1^2}{2}, \quad \tilde \Psi_2\propto x_1y_1, \quad \tilde \Psi_3 \propto y_1z_1.
\end{equation}
Equation~\eqref{5:110} clearly shows that SHG in this geometry is forbidden at $B=0$: The state $\tilde \Psi_3$ is quadrupole active ($\mathbf k\parallel z_1$, $\mathbf P\parallel y_1$) but cannot be excited by two photons polarized in the $(x_1,y_1)$ plane, while the states $\tilde \Psi_{1,2}$ are quadrupole forbidden (as they do not contain the products $z_1 x_1$ or $z_1 y_1$ which are relevant for $\mathbf k\parallel z_1$).

The magnetic field activates SHG. Due to the Zeeman effect, the field mixes $\tilde \Psi_3$ with a two-photon active exciton state. Then the second harmonic is generated via two-photon dipole excitation and quadrupolar one-photon emission. Among the three states in Eq.~\eqref{5:110} the state $\tilde \Psi_2$ is unaffected to first order by the Zeeman interaction for $\mathbf B \parallel x_1 \parallel [110]$ (it is mixed with the $\Gamma_3^+$ exciton which is far away in energy), while the states $\tilde \Psi_1$ and $\tilde \Psi_3$ are mixed into the linear combinations
\begin{equation}
	\label{Z:comb}
	\tilde \Psi_\pm = \frac{\tilde \Psi_1 \pm \mathrm i\Psi_3}{\sqrt{2}}.
\end{equation}
Each of the superposition states is simultaneously active in the two-photon excitation and in the quadrupolar emission. Both states provide a contribution to the polarization at double frequency of the same absolute value but of different signs:
\begin{widetext}
	\begin{equation}
		\label{P2omega:Z}
		P_{y_1} \propto E_{x_1}^2 \left[\frac{M_Q^* M_{TPA}}{2\hbar \omega - E_g- E_{nS}-\Delta_B/2 +\mathrm i \Gamma_{nS}} - \frac{M_Q^* M_{TPA}}{2\hbar \omega - E_g-E_{nS}+\Delta_B/2 +\mathrm i \Gamma_{nS}}\right] \approx E_{x_1}^2 \Delta_B\frac{M_Q^* M_{TPA}}{(2\hbar \omega - E_g -E_{nS} +\mathrm i \Gamma_{nS})^2}.
	\end{equation}
\end{widetext}
Here we focus on the susceptibility in the vicinity of a given $nS$-exciton resonance, $2\hbar\omega \approx E_g+ E_{nS}$, and, to shorten we introduce the following notations: $M_Q\equiv M_Q(n)\propto k_{z_1}$ is the quadrupolar transition matrix element~\eqref{M:Q:ns:5}, $M_{TPA}\equiv M_{TPA}(n)$ is the two-photon matrix element \eqref{M:TPA:ns:5} and $\Delta_B = g_X \mu_B B_{y_1}$ is the Zeeman splitting of the exciton with $g_X$ being the exciton $g$-factor and $\mu_B$ being the Bohr magneton. The second approximate equality is valid for weak Zeeman splitting $|\Delta_B| \ll \Gamma_{nS}$. This mechanism contributes to the components of the susceptibility $\chi_{y_1z_1x_1x_1x_1}$. In weak fields the polarization at the double frequency grows linearly in $B$, while for larger fields $|\Delta_B| \gg \Gamma$ the lines corresponding to the $\tilde \Psi_{\pm}$ states are significantly split and the SHG enhancement with the field becomes weaker, mainly, due to the diamagnetic effect~\cite{semina:18}.

We turn now to the magneto-Stark mechanism where the two-photon active $S$-shell exciton is mixed with the $P$-shell exciton via the equivalent electric field $\mathbf E_{\text{MSE}} \propto [\mathbf k \times \mathbf B]$. In our geometry this electric field is directed along the $y_1$ axis. The state $\tilde \Psi_1$ remains unaffected by the MSE to first order in $\mathbf B$. The state $\tilde \Psi_2$ which is not active in the Zeeman mechanism is mixed with the $P$-shell exciton and produces a double-frequency polarization along the $x_1$-axis:
\begin{equation}
	\label{P2omega:1:MS:2}
	P_{x_1} \propto E_{x_1}E_{y_1} \frac{\Delta_{\rm MSE}}{\Delta_{SP}} \frac{M_D^* M_{TPA}}{2\hbar \omega - E_g - E_{nS}+\mathrm i \Gamma_{nS}},
\end{equation}
contributing to the susceptibility $\chi_{x_1z_1x_1x_1y_1} = \chi_{x_1z_1x_1y_1x_1}$. Here
\begin{equation}
	\label{Delta:MSE}
	\Delta_{\rm MSE}=\frac{e\hbar}{Mc}k_{z_1}B_{y_1}\langle\Phi_{nS}|x_1|\Phi_{nP}\rangle,
\end{equation}
is the magneto-Stark mixing parameter, $M$ is the exciton translational motion mass, $\Delta_{SP}$ is the splitting between the nearest $S$ and $P$ exciton states (in the quasi-resonant approximation we consider only the nearest states), and $M_D$ is the matrix element of the dipole emission from the $P$-shell excitons. Note that for the $\Gamma_5^+$ $D$-shell excitons the result is similar. Also, as mentioned before, generally the $S$-$D$ mixing of the $\Gamma_5^+$ states should be taken into account.

It is instructive to estimate the relative efficiencies of the Zeeman and magneto-Stark effects for the SHG activation. We consider the weak magnetic field regime with $|\Delta_B| \ll \Gamma$ where the ratio of the corresponding contributions to the susceptibilities can be approximated as
\begin{equation}
	\left|\frac{\chi_{\rm Z}}{\chi_{\rm MSE}}\right|\sim\left|\frac{\Delta_B}{\Delta_{\rm MSE}}\frac{M_Q}{M_D}\frac{\Delta_{SP}}{\Gamma_{nS}}\right|.
\end{equation}
For rough estimates we take $g_X = 2$, disregard the difference between the exciton translational mass, the reduced mass of the electron-hole pair and the free electron mass and use Eq.~\eqref{M1:toy} to evaluate the ratio
\begin{equation}
	\frac{M_Q}{M_D}\sim \frac{q \Phi_{nS}(0)}{\Phi'_{nP}(0)}.
\end{equation}
Finally we obtain for the ration of corresponding susceptibilities:
\begin{equation}
	\left|\frac{\chi_{\rm Z}}{\chi_{\rm MSE}}\right|\sim \left|\frac{1}{\langle\Phi_{nS}|x_1|\Phi_{nP}\rangle}\frac{\Phi_{nS}(0)}{\Phi'_{nP}(0)}\frac{\Delta_{SP}}{\Gamma_{nS}}\right|.
\end{equation}
For small principal quantum numbers $n=1,2,3$ the combination of the wavefunctions gives a numerical factor on the order of unity and $\Delta_{SP} \gg \Gamma$ (the fine structure splitting between the different shells belonging to a particular multiplet $n$ is well resolved in the experiment). Thus, for low energy excitons the Zeeman effect should be dominant. For large $n\gtrsim 5$ one can use the scaling arguments~\cite{scaling}, representing $\Delta_{SP}$ in the model of quantum defects as $\Delta_{SP} = \mathcal R \delta /n^3$, where $\mathcal R$ is the exciton Rydberg energy. Further, one can evaluate the matrix elements using hydrogenic wavefunctions and recast the S-exciton linewidth as $\Gamma_{nS}= \gamma/n^3$~\cite{Nature} so that one obtains the following approximate scaling 
\begin{equation}
	\label{scaling}
	\left|\frac{\chi_{\rm Z}}{\chi_{\rm MSE}}\right|\sim \frac{1}{n^2} \times \frac{\mathcal R \delta}{\gamma}, 
\end{equation}
meaning that the MSE contributions become progressively more important for Rydberg excitons. This can be expected, since for high $n$ excitons the dipole coupling between the $S$- and $P$-shell states becomes progressively larger.

For the same reason the magneto-Stark effect can activate $P$-shell excitons which are weak in the absence of the magnetic field, see Sec.~\ref{sec:micro:B0}. The calculation shows that the MSE contribution to SHG on the $nP$-exciton takes a form similar to Eq.~\eqref{P2omega:1:MS:2}:
\begin{equation}
	\label{P:MSE:micro}
	P_i \propto E_k E_l \frac{\Delta_{\rm MSE}}{\Delta_{SP}} \frac{M_D^* M_{TPA}}{2\hbar \omega - E_{g} - E_{nP}+\mathrm i \Gamma_{nP}}.
\end{equation}
As a result, in contrast to the zero magnetic field case, the second harmonic intensities on the $S$- and $P$-excitons due to the MSE can be comparable.

\subsubsection{$\Gamma_1^+$ and $\Gamma_3^+$ excitons}\label{sec:micro:B2}

The wavefunction of the $\Gamma_1^+$ $D$-shell exciton state transforms $\propto x^2+y^2+z^2 = x_1^2+y_1^2+z_1^2$.
As mentioned, this state does not manifest itself at $B=0$ because its polarization $\mathbf P \parallel \mathbf k$ and cannot contribute to the transversal wave. The Zeeman effect mixes this state with the $\Gamma_4^+$ state which is magneto-dipole active with an oscillating magnetic moment $\mathbf \mu \parallel \mathbf B$. As a result, in a magnetic field the $\Gamma_1^+$ exciton becomes active in the polarization $\mathbf P \parallel [\mathbf k\times {\bm \mu}] \parallel [\mathbf k\times \mathbf B]$. In our geometry with $\mathbf B\parallel x_1$ and $\mathbf k\parallel z_1$ this corresponds to $\mathbf P \parallel y_1$. Hence, this state contributes to the susceptibility component $\chi_{y_1z_1x_1 x_1x_1} = \chi_{y_1z_1x_1 y_1y_1}$. The corresponding contribution to the polarization can be readily evaluated as
\begin{equation}
	\label{P:Z:gamma1}
	P_{y_1} \propto |E|^2 \frac{\Delta_B'}{\Delta_{14}} \frac{M_B^* M_{TPA}'}{2\hbar\omega - E_g - E_{nD1} + \mathrm i \Gamma_{nD1}},
\end{equation}
where $E_{nD1}$ and $\Gamma_{nD1}$ denote the energy and damping of the corresponding $D$-shell exciton, $M_B$ is the matrix element of the magnetic-dipole transition and $M_{TPA}'$ is the matrix element of the two-photon excitation of the $\Gamma_1^+$ state, $\Delta_B'$ is the Zeeman splitting and $\Delta_{14}$ is the energy separation from the nearest $\Gamma_4^+$ $D$-shell state. In addition to the Zeeman effect, the $\Gamma_1^+$ state is mixed by the magneto-Stark effect with the $P_{y_1}$ state giving rise to 
\begin{equation}
	\label{P:MS:gamma1}
	P_{y_1} \propto |E|^2 \frac{\Delta_{\rm MSE}'}{\Delta_{DP}}\frac{M_D^* M_{TPA}'}{2\hbar\omega - E_g - E_{nD1} + \mathrm i \Gamma_{nD1}}.
\end{equation}
Here $\Delta_{DP}$ is the splitting between the $D$-shell and $P$-shell states, $\Delta_{\rm MSE}'$ is the magneto-Stark parameter defined similarly to Eq.~\eqref{Delta:MSE}, but for the $D$-shell states. The comparison of Eqs.~\eqref{P:Z:gamma1} and \eqref{P:MS:gamma1} shows that the MSE is likely to dominate the second harmonic generation: Indeed, both $\Delta_{14}$ and $\Delta_{DP}$ are determined by the quantum defects and are, generally, of the same order of magnitude, while the ratio of quantities in the first fraction is
\[
\left|\frac{\Delta_B' M_B}{\Delta_{\rm MSE}'M_D} \right| \sim \left|\frac{a_0^2 \Phi_{nD}''(0)}{\langle \Phi_{nD} |x_1| \Phi_{nP}\rangle \Phi_{nP}'(0)}\right| \ll 1,
\]
see the discussion at the end of Sec.~\ref{sec:micro:B0}.

Similar mechanisms can activate the $\Gamma_3^+$ states. In the $x_1$, $y_1$ and $z_1$ system of axes the wavefunctions of the doublet read
\begin{equation}
	\label{3:110}
	\tilde \Phi_1 = 2y_1^2 - x_1^2 - z_1^2, \quad \tilde \Phi_2 = \frac{\sqrt{3}}{2}x_1z_1.
\end{equation}
For illustration we calculate the contribution via the magneto-Stark effect to the SHG, taking into account mixing of these states with the $\Gamma_4^-$ $P$-shell excitons. The state $\tilde \Phi_2$ does not play a role, while the state $\tilde \Phi_1$ provides the contribution
\begin{equation}
	\label{P2omega:3:MS}
	P_{y_1} \propto (2E_{y_1}^2 - E_{x_1}^2) \frac{\tilde \Delta_{\rm MSE}}{\tilde \Delta_{DP}}\frac{M_D^* M_{TPA}}{2\omega - E_g - E_{nD3} + \mathrm i \Gamma_{nD3}},
\end{equation}
where $\tilde \Delta_{\rm MSE}$ and $\tilde \Delta_{DP}$ are the corresponding mixing parameter and the separation from the nearest $P$-state, respectively, $E_{nD3}$ and $\Gamma_{nD3}$ are the energy and damping of the $D$-shell $\Gamma_3^+$ exciton.

To summarize the microscopic theory, we have identified the main mechanisms and the intermediate states for SHG on the odd and even excitons in Cu$_2$O. We have demonstrated that at $B=0$ the $S/D$ excitons of $\Gamma_5^+$ symmetry provide the dominant contribution to SHG, while the $P$-excitons provide a parametrically smaller contribution, see Eq.~\eqref{rat:SP:1}. The $D$-excitons of $\Gamma_3^+$ symmetry provide contributions which are smaller than that of the $P$-excitons and of the $S/D$-excitons due to the necessity of involving transitions via very distant bands. In the presence of a magnetic field, we have identified two main SHG mechanisms, the Zeeman effect and the magneto-Stark effect, and demonstrated that with increasing exciton principal quantum number the MSE contribution dominates. Also, the MSE can provide similar strengths of the $P$ and $S$ excitons in the SHG effect.

\section{EXPERIMENT}
\label{sec:experiment}
\begin{figure}[h]
	\begin{center}
		\includegraphics[width=0.5\textwidth]{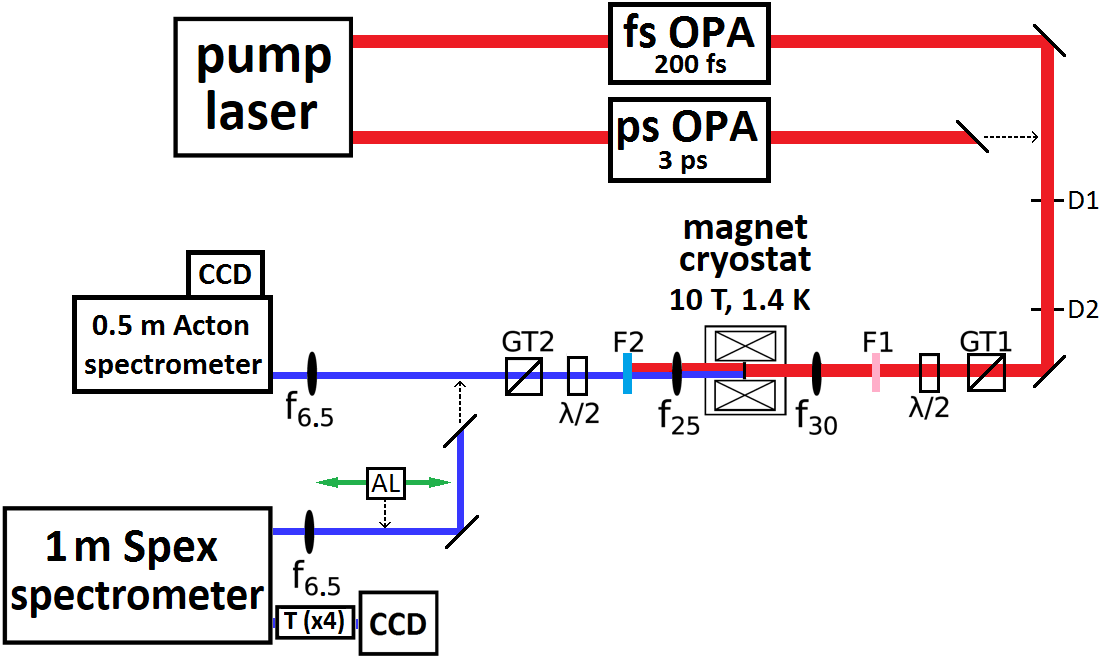}
		\caption[Scan]{Setup for SHG spectroscopy: AL - alignment laser, CCD - charge-coupled device camera, D - diaphragm, F - color filter, f$_\textrm{xx}$ - lens with xx-cm focal length, GT - Glan Thompson linear polarizer, $\lambda/2$ - half-wave plate, OPA - optical parametric amplifier, T($\times4$) - telescope with a magnification factor of four. The double side alignment laser (AL) in front of the $\SI{1}{\metre}$ Spex spectrometer is useful for accurate aligning the SHG beam into the Spex spectrometer.}
		\label{fig:experiment}
	\end{center}
\end{figure}

Our experimental setup is similar to the setup described in Ref.~\cite{Mund1}. As shown in Fig.~\ref{fig:experiment}, we have now the choice between two detection systems: (i) an $\SI{0.5}{\metre}$ Acton spectrometer ($5\times5\;\si{\centi\metre}^2$-sized grating with $1800$ grooves/$\si{\milli\metre}$ in first order) connected to a CCD camera ($400\times1340$ pixel of size $\SI{20}{\micro\metre}$), leading to a spectral resolution of $\SI{80}{\micro\electronvolt}$ around $\SI{2}{\electronvolt}$ photon energy; (ii) a $\SI{1}{\metre}$ Spex spectrometer ($10\times10\;\si{\centi\metre}^2$ sized grating with $1200$ grooves/$\si{\milli\metre}$, used in first or second order), combined with a $4\times$ amplification on the detection CCD camera ($512\times2048$ pixel of size $\SI{13.5}{\micro\metre}$) leading to a resolution of $\SI{20}{\micro\electronvolt}$ in first order and $\SI{10}{\micro\electronvolt}$ in second order. 

\begin{figure}[h]
	\begin{center}
		\includegraphics[width=0.48\textwidth]{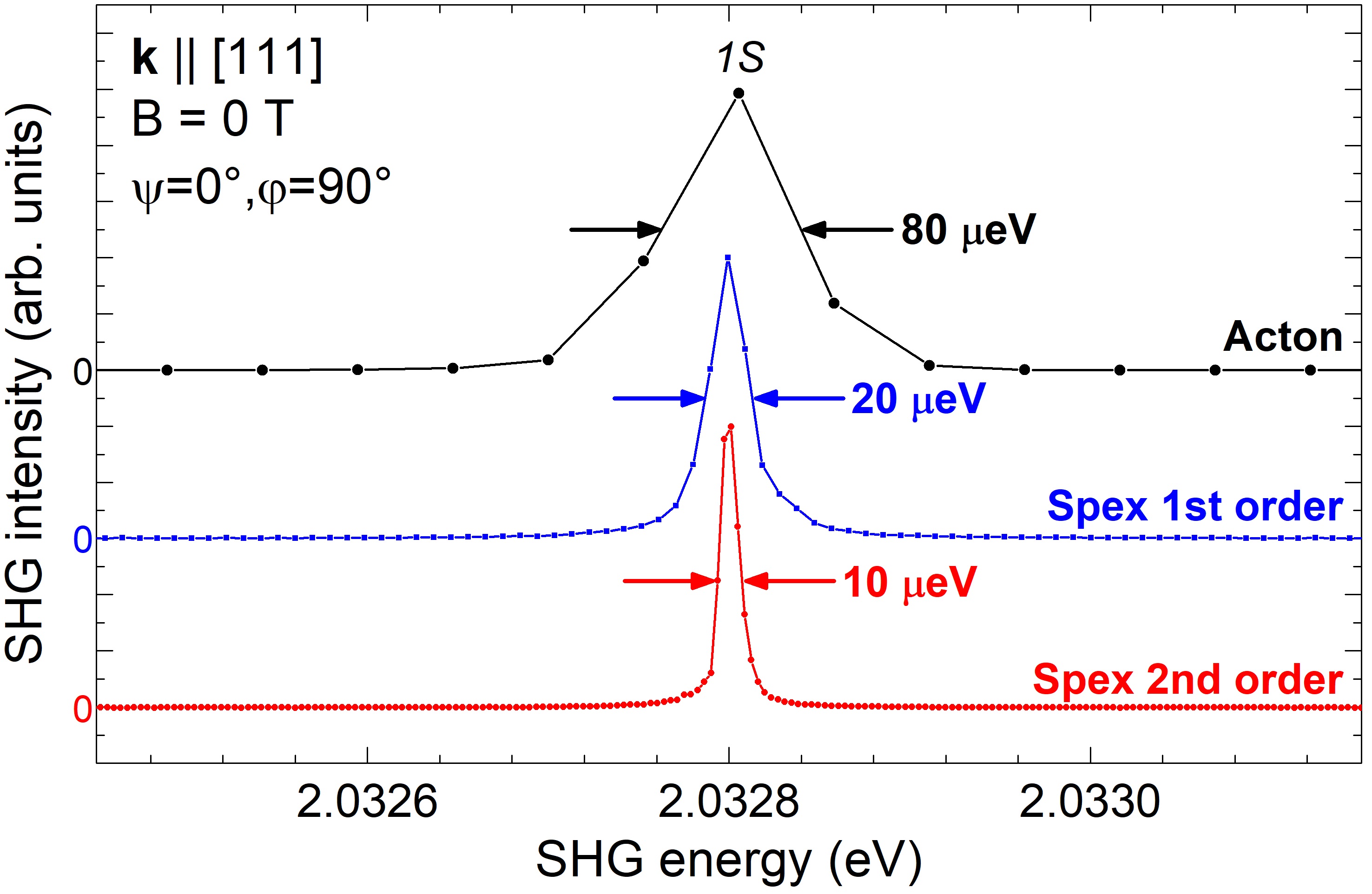}
		\caption[Scan]{Crystallographic SHG spectra of the $1S$ orthoexciton ($\mathbf k \parallel [111]$, $x \parallel \mathbf E^\omega\parallel [1\bar10]$ and $y \parallel \mathbf E^{2\omega} \parallel [11\bar2]$) excited by femtosecond pulses at a temperature of $\SI{1.4}{\kelvin}$ for demonstration of the resolution of the two spectrometers.}
		\label{fig:resolution}
	\end{center}
\end{figure}

In Figure~\ref{fig:resolution} we present SHG spectra recorded in the spectral range of the $1S$ orthoexciton at zero magnetic field. Light propagation along the $[111]$-direction is chosen, making SHG possible also without application of a magnetic field. The data allow us to compare the resolution for the two detection systems (Acton spectrometer and Spex spectrometer used in different orders). The Spex spectra confirm that the larger focal length in combination with the implementation of the $4\times$ magnification optics in front of the CCD camera helps to improve the spectral resolution significantly, in particular in 2nd order. 

The laser system (Light Conversion) provides femto- or picosecond pulses with a repetition rate up to $\SI{30}{\kilo\hertz}$. For the SHG experiments we use the femtosecond pulses with a duration of $\SI{200}{\femto\second}$, corresponding to a spectral full width at half maximum (FWHM) of $\SI{10}{\milli\electronvolt}$. For the excitation of excitons with $n \geqslant 3$ the laser is set to $\SI{1.08}{\electronvolt}$, using an average power of $\SI{20}{\milli\watt}$. For measuring the two-photon absorption using photoluminescence excitation (TP-PLE) experiments we tune the ps laser through the resonances and monitor the TPA through the emission from the $1S$ exciton and/or its $\Gamma^-_3$ phonon side band. The spectral resolution of the TPA spectra is limited by the spectral width of the $\SI{3.3}{\pico\second}$ pulses to $\SI{0.7}{\milli\electronvolt}$. The laser beam was focused on the sample to a spot with a diameter of $\SI{100}{\micro\meter}$. At an average power of $\SI{20}{\milli\watt}$ the laser intensity on the sample surface is $\SI{2.5}{\giga\watt\per\centi\meter\squared} $.

The samples are cut from a natural Cu$_2$O crystal in different crystalline orientations and thicknesses. The samples are mounted strain-free in a split-coil superconducting magnet allowing a magnetic field strength up to $\SI{10}{\tesla}$ at a sample temperature as low as $\SI{1.4}{\kelvin}$. The polarization angles of the ingoing laser beam ($\psi$) and the SHG light ($\varphi$), see Fig.~\ref{fig:orientation}, can be tuned independently by automatized polarizers controlled with a LabVIEW program. It should be noted, that the $180^\circ$ periodicity of all results is expected since a phase shift by $180^\circ$ in the amplitudes (experimentally setting the $\lambda/2$ plates) has no influence on the $1$D and $2$D SHG intensity plots. For convenience the angular dependences are mostly taken only in the range from $0^\circ$ to $180^\circ$ and then the same data are extended to the range $180^\circ$ to $360^\circ$. It was proved that tuning the $\lambda/2$ plates through the whole range $0^\circ$ to $360^\circ$ did not lead to any novel deviating information. For the $2$D plots the polarizer angle $\psi$ is rotated in steps of $10^\circ$ for the full rotation starting at an analyzer angle $\varphi$ of $0^\circ$. This is repeated for the analyzer angle $\varphi$ in steps of $10^\circ$ for the full rotation. It takes four hours to measure the full polarization dependence. For the $1$D plots the polarization angles are varied in steps of $5^\circ$

\section{Experimental results and discussion}
\label{sec:experimental results}
	
\begin{figure}[h]
	\begin{center}
		\includegraphics[width=0.48\textwidth]{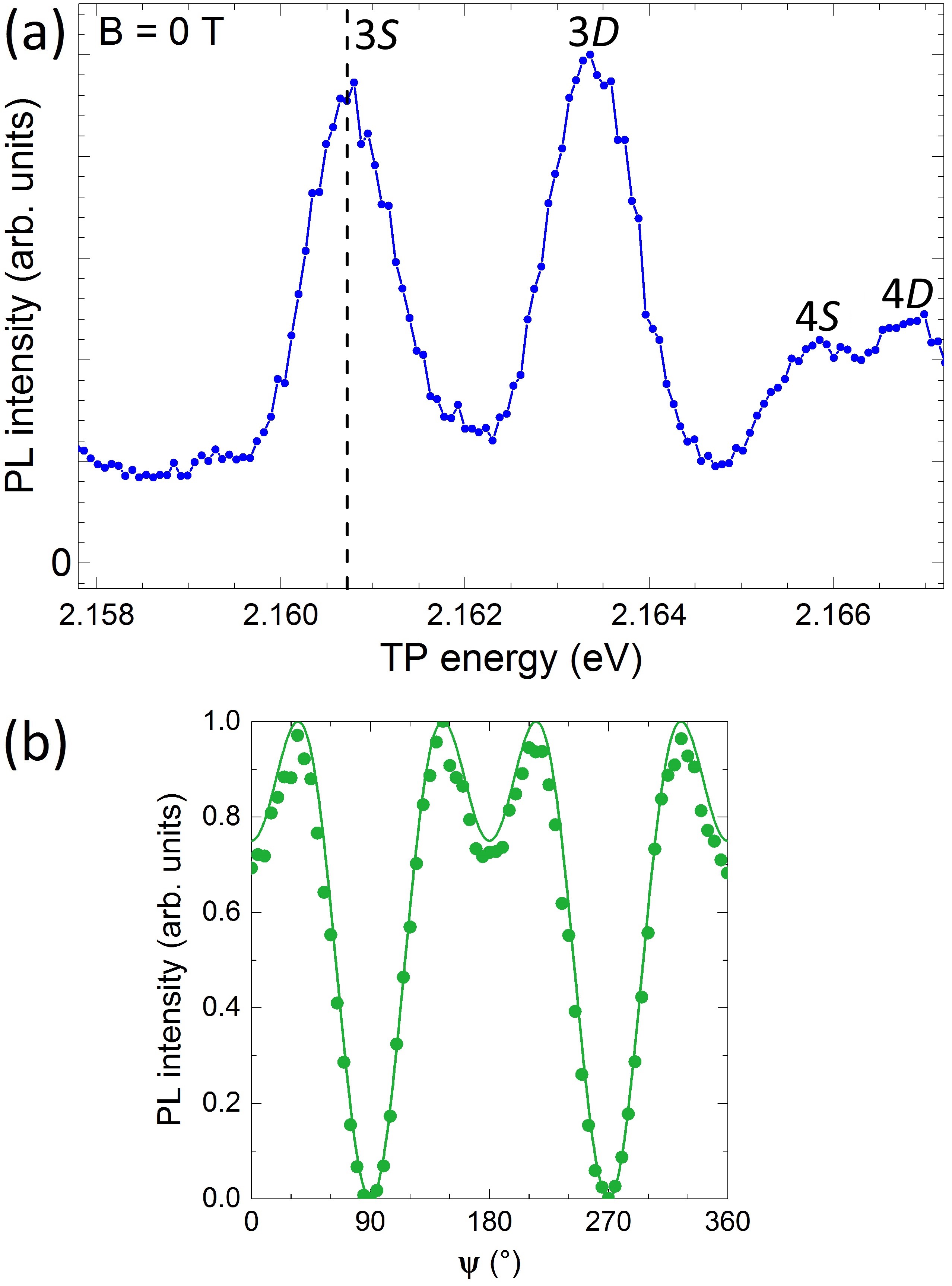}
		\caption[Scan]{(a) Zero-magnetic field PLE spectrum for TPA of the the $n=3$ and $n=4$ excitons. (a) The spectrum is recorded by scanning the laser in the $\si{\pico\second}$ configuration and detecting the luminescence from the $1S$ exciton or its $\Gamma_3^-$ phonon replica. The dashed line ($\SI{2.167}{\electronvolt}$) marks the spectral position of the $3S$ exciton. (b) Measured (dots) and simulated (solid line) [Eq.~\eqref{eqn:ITPA}] TPA polarization dependence of the $3S$ exciton on the polarization angle $\psi$ of the ingoing laser. The sample is oriented as shown in Fig.~\ref{fig:orientation} ($\textbf{k}\parallel[1\bar 1 0]$, $x\parallel\mathbf B\parallel[110]$ and $y\parallel[001]$).}
		\label{fig:TPAspectrum}
	\end{center}
\end{figure}

In this section we will present experimental data for nonlinear optical effects with the main emphasis on SHG from the $S$ and $D$ excitons with $n\geqslant 3$ using the configuration $\mathbf k \parallel [1\bar1 0], \mathbf B \parallel [110]$ and thus $\mathbf E_{\text{MSE}} \parallel [001]$. 
In the zero-field case SHG is forbidden, but TPA is allowed. In Figure~\ref{fig:TPAspectrum}(a) we show the TPA spectrum of $n = 3$ and $n = 4$ exciton multiplets. For each multiplet we observe two features of similar intensity and linewidth (determined by the width of the laser pulses). The energies of the lines and the splitting between them are in good agreement with previous measurements~\cite{Uihlein,Mund1}, where we have assigned them to excitons with dominant $S$- and $D$-envelope with the $\Gamma_5^+$ symmetry of the total exciton wavefunction. In accordance with the microscopic theory, the TPA-excitation of the $D$-excitons becomes allowed mainly due to mixing with the $S$-excitons.

In Figure~\ref{fig:TPAspectrum}(b) we present the TPA polarization dependence of the $3S$ exciton (dots) as function of the linear polarization $\psi$ of the ingoing laser. The experimental data agree well with the expected dependence (solid line)
\begin{equation}
\label{TPA:Gamma5}
I_{\text{TPA}}(\psi) \propto \frac{1}{2}(\cos^4{\psi}+\sin^2{2\psi}),
\end{equation}
derived from our symmetry analysis in Sec. \ref{shg polarization dependences}, Eq.~\eqref{eqn:ITPA}, see also Eq.~\eqref{5:110}.

\begin{figure}[h]
	\begin{center}
		\includegraphics[width=0.48\textwidth]{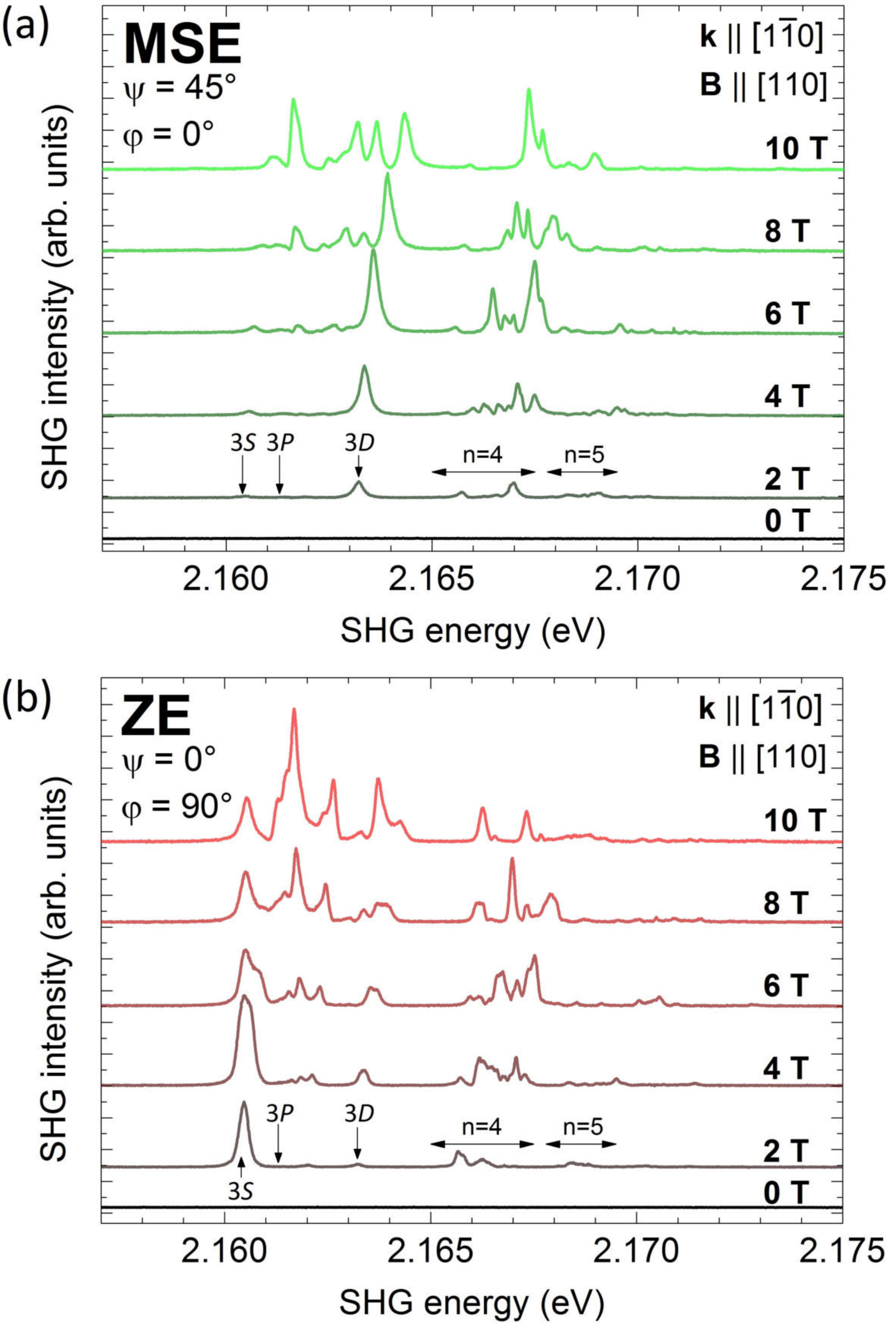}
		\caption[Scan]{Magnetic field-dependent SHG spectra induced by the magneto-Stark effect (a) or the Zeeman effect (b) in the energy range starting from the $n = 3$ excitons, where features related to the $n = 4$ and $5$ multiplets are observed. The central photon energy of the fs-pulsed fundamental laser is set to $\SI{1.082}{\electronvolt}$, see also the spectrum in Fig.~\ref{fig:gamma31}. The sample is oriented as shown in Fig.~\ref{fig:orientation} ($\textbf{k}\parallel[1\bar 1 0]$, $x\parallel\mathbf B\parallel[110]$ and $y\parallel[001]$). The polarizations $\psi$ and $\varphi$ are chosen according to Fig.~\ref{fig:gamma5}, which allows a distinction between MSE ($\psi=45^{\circ}$, $\varphi=0^{\circ}$) and ZE ($\psi=0^{\circ}$, $\varphi=90^{\circ}$).}
		\label{fig:Bserie}
	\end{center}
\end{figure}

Now we turn to the analysis of the SHG process in a magnetic field. A basic result of the corresponding analysis, see Sec.~\ref{sec:dom}, is that the configuration $\mathbf k \parallel [1\bar10]$ and $\mathbf B \parallel [110]$ allows one to distinguish between SHG induced by the ZE and the MSE and further allows identification of weaker processes associated with the $\Gamma_1^+$ and $\Gamma_3^+$ exciton states, see Sec.~\ref{sec:weak}.
Figure \ref{fig:Bserie} shows SHG spectra for increasing magnetic field from $0$ up to 10~T, where we have chosen the polarization configurations that are supposed to allow distinction between MSE-induced SHG (a) and ZE-induced SHG (b) as indicated in Fig.~\ref{fig:gamma5}(a) (MSE) and Fig.~\ref{fig:gamma5}(b) (ZE). The spectra show the energy range starting from $n = 3$. Besides $n = 3$ lines also features associated with $n = 4$ and $n = 5$ are seen. As expected, the SHG is only magnetic-field-induced for the chosen configuration, and one also sees a strong overall enhancement of the SHG intensity with increasing magnetic field. Simultaneously, there are striking differences in the appearance of the spectra for the two configurations: different spectral lines and strong intensity variations show up. For example, for the ZE-related SHG most intensity occurs on the low energy flank of the $n =3$ multiplet, while for the MSE-related SHG the intensity is shifted towards the high energy flank.

\begin{figure}[h]
	\begin{center}
		\includegraphics[width=0.5\textwidth]{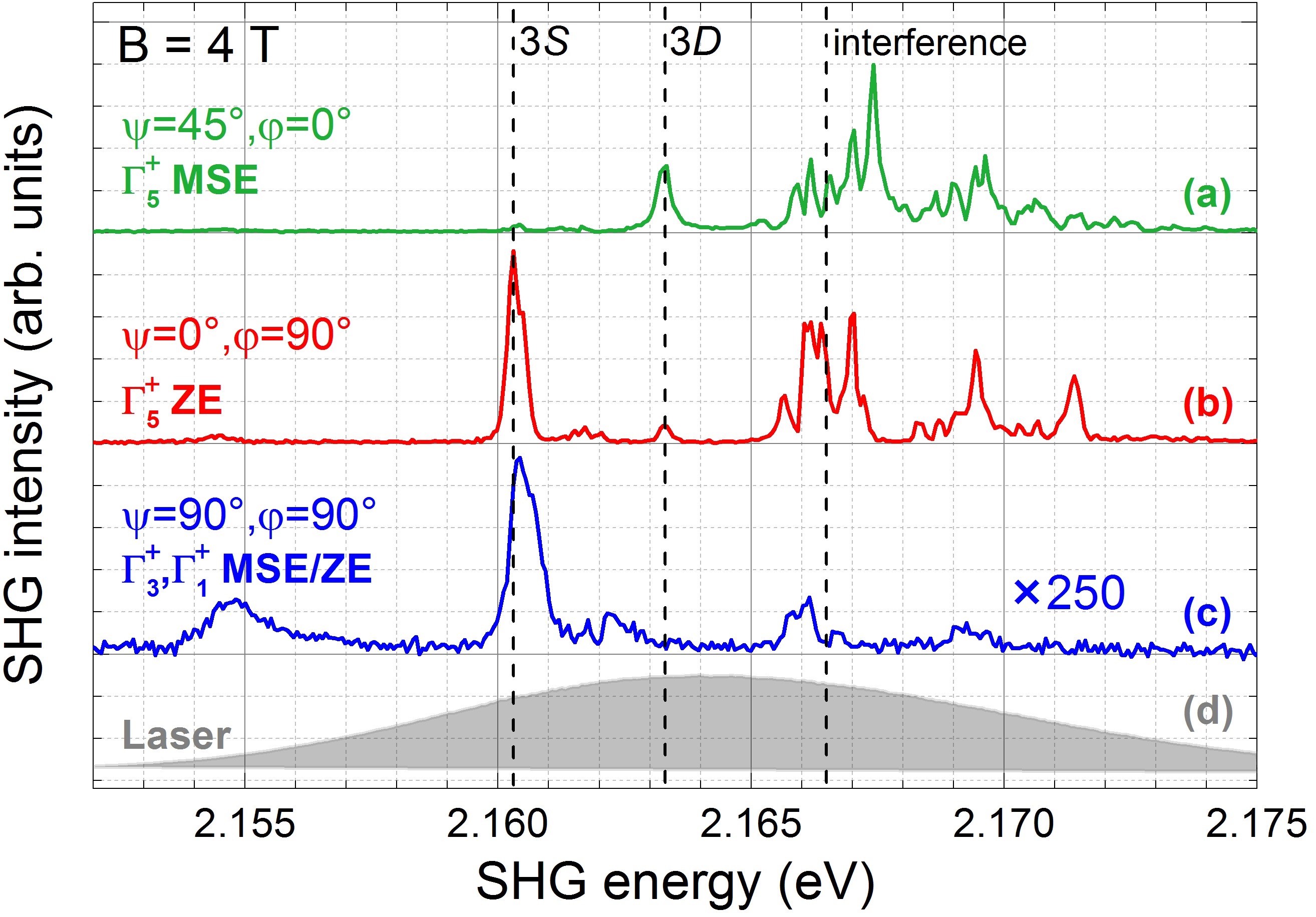}
		\caption[Scan]{SHG at $\SI{4}{\tesla}$ for selected polarization configurations. (a) MSE spectrum and (b) ZE spectrum correspond to the spectra for $\SI{4}{\tesla}$ in Fig.~\ref{fig:Bserie}. In (c) a SHG spectrum of the \textit{Weaker Processes} [Sec.~ \ref{sec:weak}] which are due to $\Gamma_1^+$ and $\Gamma_3^+$ $D$-states are shown, which are distinguished from the strong resonances (about a factor 250) of (a) and (b) by the choice of the polarization configuration ($\psi =90^\circ$ and $\varphi =90^\circ$). In (d) we show SHG of the laser set to $\SI{1.082}{\electronvolt}$ as measured with BBO (beta barium borate). The sample is oriented as shown in Fig.~\ref{fig:orientation} ($\textbf{k}\parallel[1\bar 1 0]$, $x\parallel\mathbf B\parallel[110]$ and $y\parallel[001]$).}
		\label{fig:gamma31}
	\end{center}
\end{figure}

For the analysis of the data in more detail, we show in Fig.~\ref{fig:gamma31} SHG spectra at $\SI{4}{\tesla}$ and the laser spectrum. Besides the two configurations separating the ZE and the MSE, also another configuration, in which the $\Gamma_1^+$ and $\Gamma_3^+$ states are expected to contribute exclusively to the SHG with, however, comparatively weak intensity, is shown. From the ZE and MSE spectra one indeed sees the complementarity of the SHG lines in the two configurations and therefore of the underlying mechanisms for the $n = 3$ multiplet: While ZE-induced SHG appears mostly on the $3S$-exciton, the MSE-induced SHG is concentrated on the $3D$-exciton. For the multiplets with high principal quantum number the SHG spectrum becomes increasingly complex due to the multitude of involved states. For a detailed analysis of the magnetic field dependence we refer to Ref.~\cite{RommelSHG}.

Before proceeding with a detailed symmetry analysis let us briefly compare the results with the microscopic theory developed in Sec.~\ref{sec:micro} in terms of the states providing stronger and weaker contributions to the ZE-induced and MSE-induced SHG. It follows from Fig.~\ref{fig:Bserie}(b) that for the ZE the intensities of the $P$-excitons in moderate fields ($2\ldots 4$~T) are considerably smaller than those for the $S$ and even $D$ states. This is in line with the microscopic analysis showing that at $B=0$ the SHG-allowed $P$ states provide much weaker contributions to the SHG as compared to the $S$ excitons, Eq.~\eqref{rat:SP:1}. Note that the Zeeman mechanism does not mix states of different parity, it makes the otherwise forbidden $P$-excitons allowed only by rotating the microscopic dipole moment or by breaking the destructive interference of the states that are degenerate at $B=0$, cf. Eq.~\eqref{P2omega:Z}. In contrast, the MSE efficiently mixes $S$- and $P$-excitons and can result in comparable contributions of the $S$- and $P$-states to SHG, see Eq.~\eqref{P:MSE:micro}. Thus, already at moderate fields the $S$ and $P$ excitons provide similar contributions to the MSE-induced SHG, see Fig.~\ref{fig:Bserie}(a).

\begin{figure}[h]
	\begin{center}
		\includegraphics[width=0.4\textwidth]{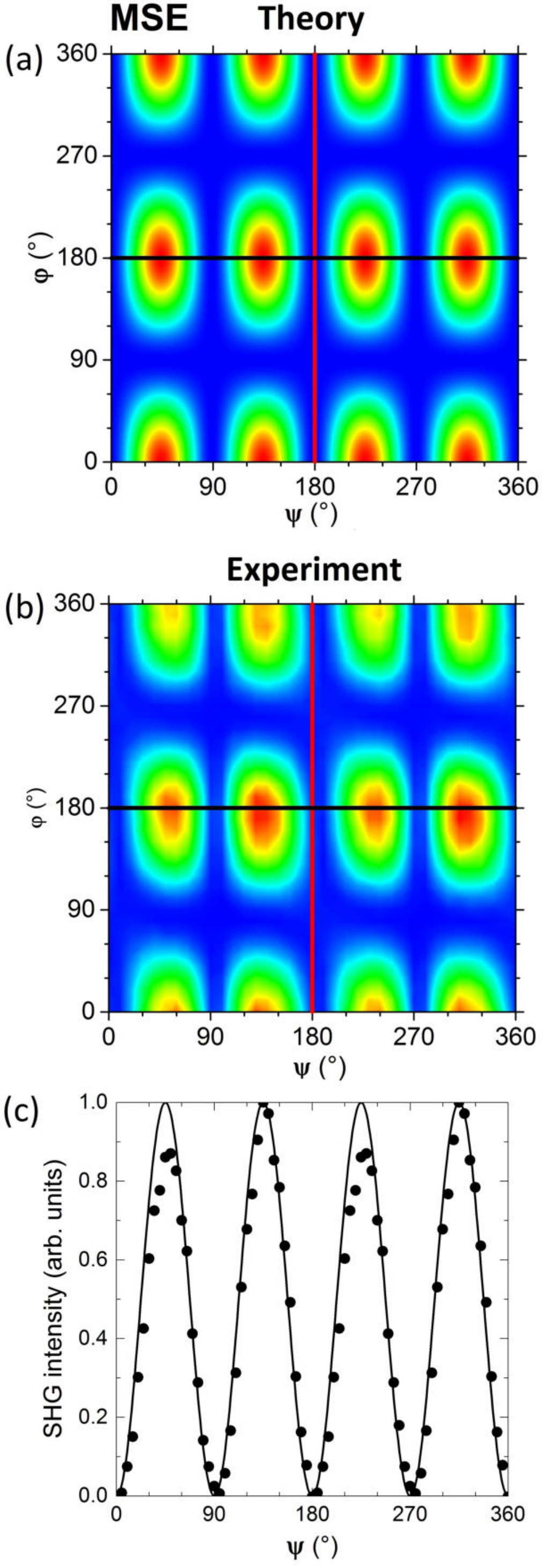}
		\caption[Scan]{Theory (a) and experiment (b) of the $2$D polarization dependence of magneto-Stark effect induced SHG intensity at the $3D$ exciton resonance ($\SI{2.1633}{\electronvolt}$) at $B=\SI{4}{\tesla}$. The theoretical results are calculated according to Eq.~\eqref{eqn:IE}. (c)~Cut through the two contour plots in (a) and (b) for varying the ingoing polarization angle $\psi$ at fixed outgoing SHG polarization angle $\varphi = 180^{\circ}$. The measured results are given by the dots and the simulations by the solid lines. The sample is oriented as shown in Fig.~\ref{fig:orientation} ($\textbf{k}\parallel[1\bar 1 0]$, $x\parallel\mathbf B\parallel[110]$ and $y\parallel[001]$).}
		\label{fig:MSE_experiment}
	\end{center}
\end{figure}

As described above, conclusive information about the underlying SHG mechanisms may be obtained by contour plots showing the SHG intensity as function of the linear polarization angles $\psi$ and $\varphi$ of the ingoing laser and the outgoing SHG light. Let us consider first the MSE-related SHG. Figure \ref{fig:MSE_experiment}(a) shows the dependencies calculated according to Eq.~\eqref{eqn:IE}, which reveal a four-fold symmetry pattern corresponding to a period of $90^\circ$, when $\psi$ is varied from $0^\circ$ to $360^{\circ}$, and $\varphi$ is fixed. On the other hand, variation of $\varphi$ gives a two-fold pattern with a period of $180^\circ$ when keeping $\psi$ constant. This unique footprint of MSE-induced SHG is nicely confirmed by the experimental data in Fig.~\ref{fig:MSE_experiment}(b) as further detailed in Fig.~\ref{fig:MSE_experiment}(c) showing the SHG intensity as function of $\psi$ along the black tuning lines in (a) and (b). Here we singled out the $3D$ resonance at $\SI{2.1633}{\electronvolt}$ marked by the left dashed line in Fig.~\ref{fig:gamma31}, following the results of our theory. Slight deviations between theory and experiment might be caused by tiny misalignments of the chosen configuration or strain in the sample, which may lead in particular to the slight distortion of the signal relative to lines with $\psi =$ const. as discussed for the $1S$ exciton in Ref.~\cite{Mund2}. Further, an intensity drift of the exciting fs laser during the rather long angle scanning time of $4$ hours may occur. 

\begin{figure}[h]
	\begin{center}
		\includegraphics[width=0.395\textwidth]{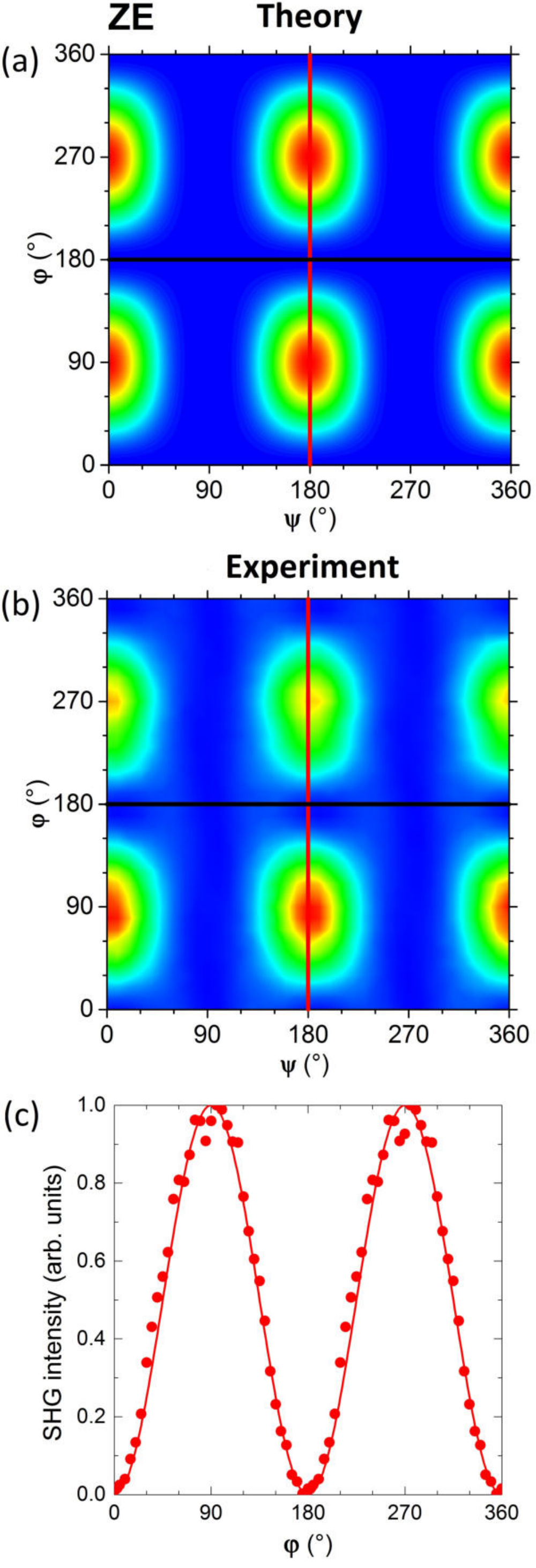}
		\caption[Scan]{Theory (a) and experiment (b) of the $2$D polarization dependence of Zeeman effect induced SHG intensity at the $3S$ exciton resonance ($\SI{2.1603}{\electronvolt}$) at $B=\SI{4}{\tesla}$. The theoretical results are calculated according to Eq.~\eqref{eqn:IB}. (c)~Cut through the two contour plots in (a) and (b) for varying the ingoing polarization angle $\psi$ at fixed outgoing SHG polarization angle $\psi = 180^{\circ}$. The measured results are given by the dots and the simulations by the solid lines. The sample is oriented as shown in Fig.~\ref{fig:orientation} ($\textbf{k}\parallel[1\bar 1 0]$, $x\parallel\mathbf B\parallel[110]$ and $y\parallel[001]$).}
		\label{fig:ZE_experiment}
	\end{center}
\end{figure}

Next we turn to the demonstration of the ZE induced SHG for which we selected the $3S$ resonance at $\SI{2.1603}{\electronvolt}$, again motivated by the symmetry analysis. The resonance is marked in Fig.~\ref{fig:gamma31} by the middle dashed line. The theoretical expectations according to the symmetry analysis are shown in the Fig.~\ref{fig:ZE_experiment}(a), visualizing Eq.~\eqref{eqn:IB}. Here twofold symmetry patterns with a period of $180^\circ$ are expected for varying one of the two basic polarizations while keeping the other constant. Also these predictions are in perfect agreement with the experimental data, see Fig.~\ref{fig:ZE_experiment}(b), confirming, e.g., the expected behaviors along the $\varphi$- and the $\psi$-axes, as detailed further in Fig.~\ref{fig:ZE_experiment}(c). Possible reasons for the slight differences between theory and experiment are the same as discussed above.

\begin{figure}[h]
	\begin{center}
		\includegraphics[width=0.4\textwidth]{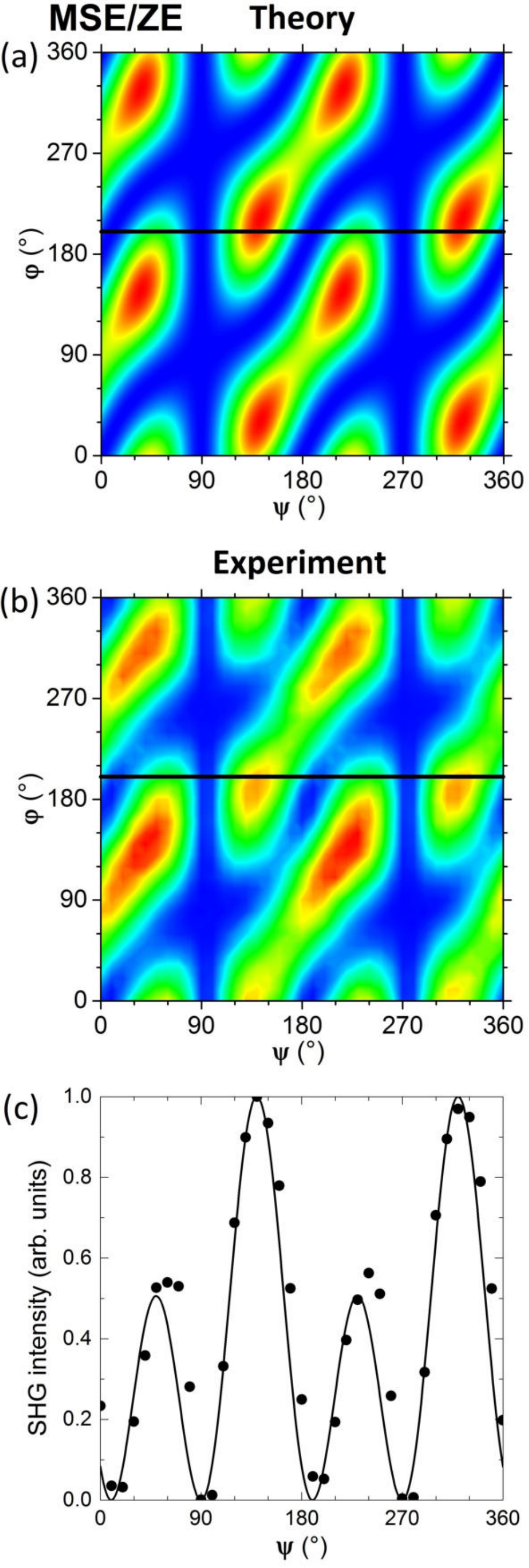}
		\caption[Scan]{Theory (a) and experiment (b) of the polarization dependent SHG intensity for the case of an interference of the magneto-Stark effect and the Zeeman effect. In the experiments we chose as detection energy $\SI{2.1664}{\electronvolt}$ at $B= \SI{4}{\tesla}$, located in the energy range of the $n = 4$-multiplet. The interference of ZE and MSE can be well described by Eq.\eqref{eqn:Itot} using the parameters $\alpha = 4/3$ and $\beta = 1$. The tuning lines in (a) and (b) indicate the 1D plot shown in (c). The sample is oriented as shown in Fig.~\ref{fig:orientation} ($\textbf{k}\parallel[1\bar 1 0]$, $x\parallel\mathbf B\parallel[110]$ and $y\parallel[001]$).}
		\label{fig:2Dinterference}
	\end{center}
\end{figure}

Since we understand now in detail the two basic origins of MSE and ZE for the magnetic-field-induced SHG signals by separating them through proper polarization configurations, we can now also assess in more detail the influence of interference effects when both of them contribute and the interference effects should be taken into consideration. According to our symmetry analysis in Sec. \ref{shg polarization dependences} the SHG intensity is given by Eq.~\eqref{eqn:Itot}, which is plotted as function of the two polarization angles in Fig.~\ref{fig:2Dinterference}(a). One immediately sees that the SHG pattern becomes distorted compared to the previous cases ($2$D plots for MSE [Fig.~\ref{fig:gamma5}(a)] and ZE [Fig.~\ref{fig:gamma5}(b)]). 
The mixing parameters $\alpha=4/3$ and $\beta = 1$ are gained from a fit of the experimental $2$D-plot shown in Fig.~\ref{fig:2Dinterference}(b). It shows the results of corresponding measurements, where we chose as energy setting $\SI{2.1664}{\electronvolt}$ in a magnetic field of $\SI{4}{\tesla}$. At this energy within the $n = 4$ multiplet we expect interference of the MSE [Fig.~\ref{fig:gamma31}(a)] and the ZE [Fig.~\ref{fig:gamma31}(b)] contributions to SHG. 
The $1$D plot in Fig.~\ref{fig:2Dinterference}(c) ($\psi$ tuning line for $\varphi = 200^\circ$ as marked in Fig.~\ref{fig:2Dinterference}(a) and \ref{fig:2Dinterference}(b)) shows again the good agreement between experiment (symbols) and theory (solid line).

To visualize the interference of both effects we present a contour plot (second animation, see Ref.~[42]), in which the relative weight of the Zeeman effect and the magneto-Stark effect interfering in the SHG generation is varied, see Eq.~\eqref{eqn:Itot}. In detail, the weight of the Zeeman effect is increased from zero to unity, corresponding to its exclusive contribution. The weight of the MSE is reduced accordingly. This situation may be obtained by adjusting the wave vector and the magnetic field properly. One clearly sees the smooth transition between these two limiting cases shown in Fig.~\ref{fig:gamma5}, by continuous distortion of the contour plots so that they transform into each other.


\FloatBarrier

\begin{figure}[htb]
	\begin{center}
		\includegraphics[width=0.365\textwidth]{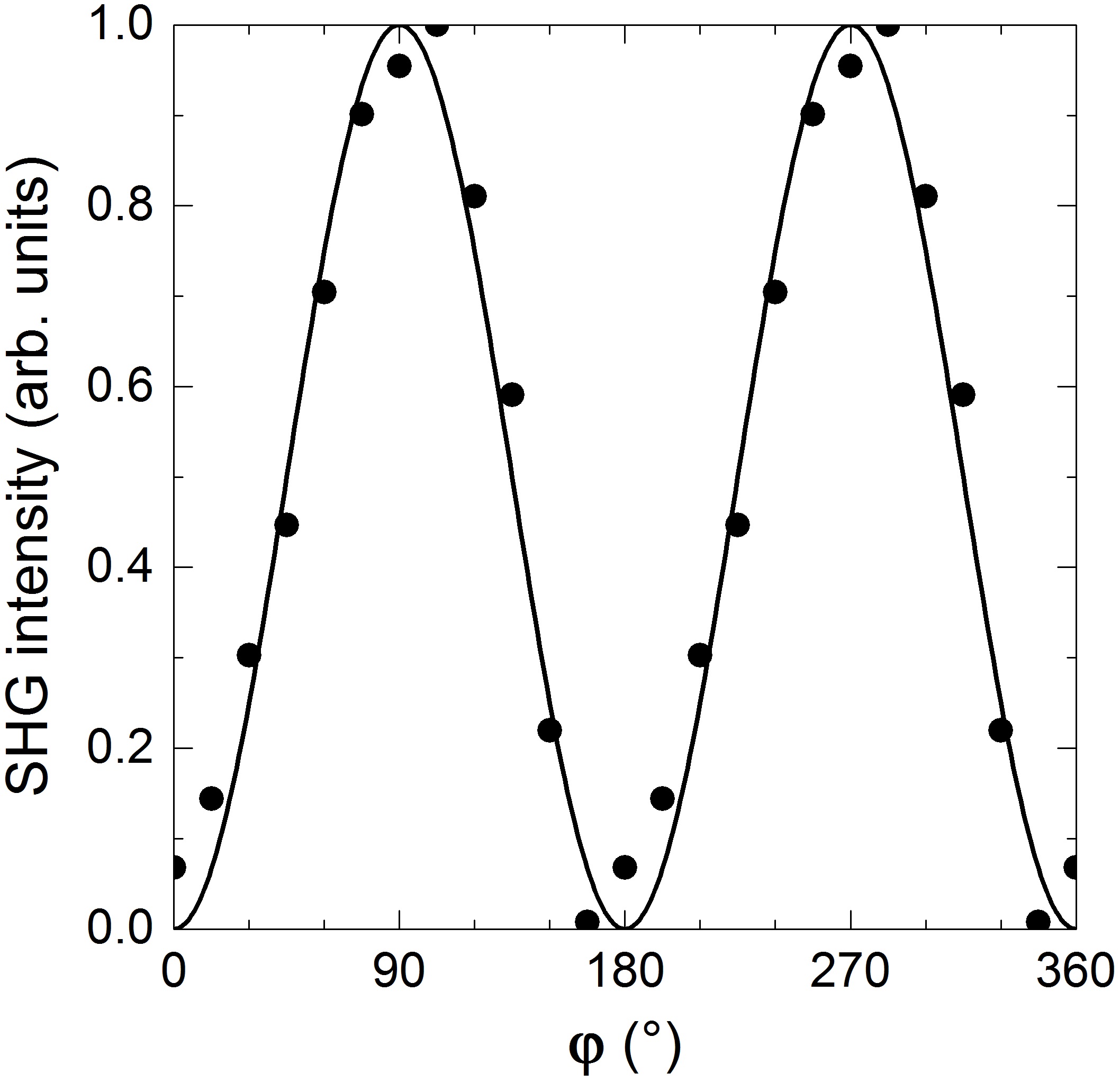}
		\caption[Scan]{Measured SHG intensity (dots) at $\SI{2.1604}{\electronvolt}$ and simulated (solid line) SHG intensity from the $\Gamma_1^+$ and $\Gamma_3^+$ $D$-states for a configuration, in which according to our symmetry analysis contributions of the $\Gamma_5^+$ excitons are suppressed (see also Fig.~\ref{fig:gamma31sim}), as function of the outgoing polarization $\varphi$ for fixed ingoing polarization $\psi=90^\circ$. The sample is oriented as shown in Fig.~\ref{fig:orientation} ($\textbf{k}\parallel[1\bar 1 0]$, $x\parallel\mathbf B\parallel[110]$ and $y\parallel[001]$).}
		\label{gamma13}
	\end{center}
\end{figure}

For the detection of weaker SHG processes related to the two-photon excitation of the $\Gamma_1^+$ and $\Gamma_3^+$ $D$-states, see subsection \ref{sec:weak} and \ref{sec:micro:B2}, we chose the configuration $\psi = 90^{\circ} = \varphi$ according to Fig.~\ref{fig:gamma31sim}, where no contributions from the stronger $\Gamma_5^+$-states to the SHG signal are expected. Doing so, we indeed observe signals [Fig.~\ref{fig:gamma31}(c)], in particular at the energies where the $\Gamma_5^+$ excitons are absent, see for example the energy range between $\SI{2.160}{\electronvolt}$ and $\SI{2.163}{\electronvolt}$. The intensity of these signals is, however, weaker by a factor of about $250$ compared to the SHG intensity level at the $\Gamma_5^+$-states (as also indicated by the increased noise level). The signal shows the expected polarization dependence, see Fig.~\ref{fig:gamma31sim}, where we have chosen a particular polarization setting as indicated by the tuning line in Fig.~\ref{fig:gamma31sim} ($\varphi$-tuning for $\psi=90^\circ$). The results along with the simulation are shown in Fig.~\ref{gamma13} revealing a characteristic two-fold symmetry pattern. Both the ZE and the MSE contribute to the SHG signal.

Having now established full agreement between the symmetry analysis and the experiment and having identified the symmetries of the excitonic states participating in the SHG as well as the particular underlying mechanisms, we turn again to the comparison of the experimental data with the microscopic theory. Namely, we address the relative intensity of the ZE and the MSE to the SHG for different principal quantum numbers. This comparison is possible by selecting two configurations in which SHG is solely induced by either the ZE and the MSE. However, developing a systematic dependence on principal quantum number is complicated by the fact, that the spectral lines from different multiplets are rather well separated only for $n = 3$ and $4$, while for higher $n$ the spectral lines of different multiplets overlap at finite magnetic fields that are, on the other hand, strong enough to obtain a reasonable SHG intensity well above the noise level. The spectral overlap of exciton features also leads to complex state mixings and anticrossings. Nevertheless, we made such an analysis up to $n = 6$, where determination of the intensities is still possible with the mentioned restrictions.

\begin{figure}[h]
	\begin{center}
		\includegraphics[width=0.45\textwidth]{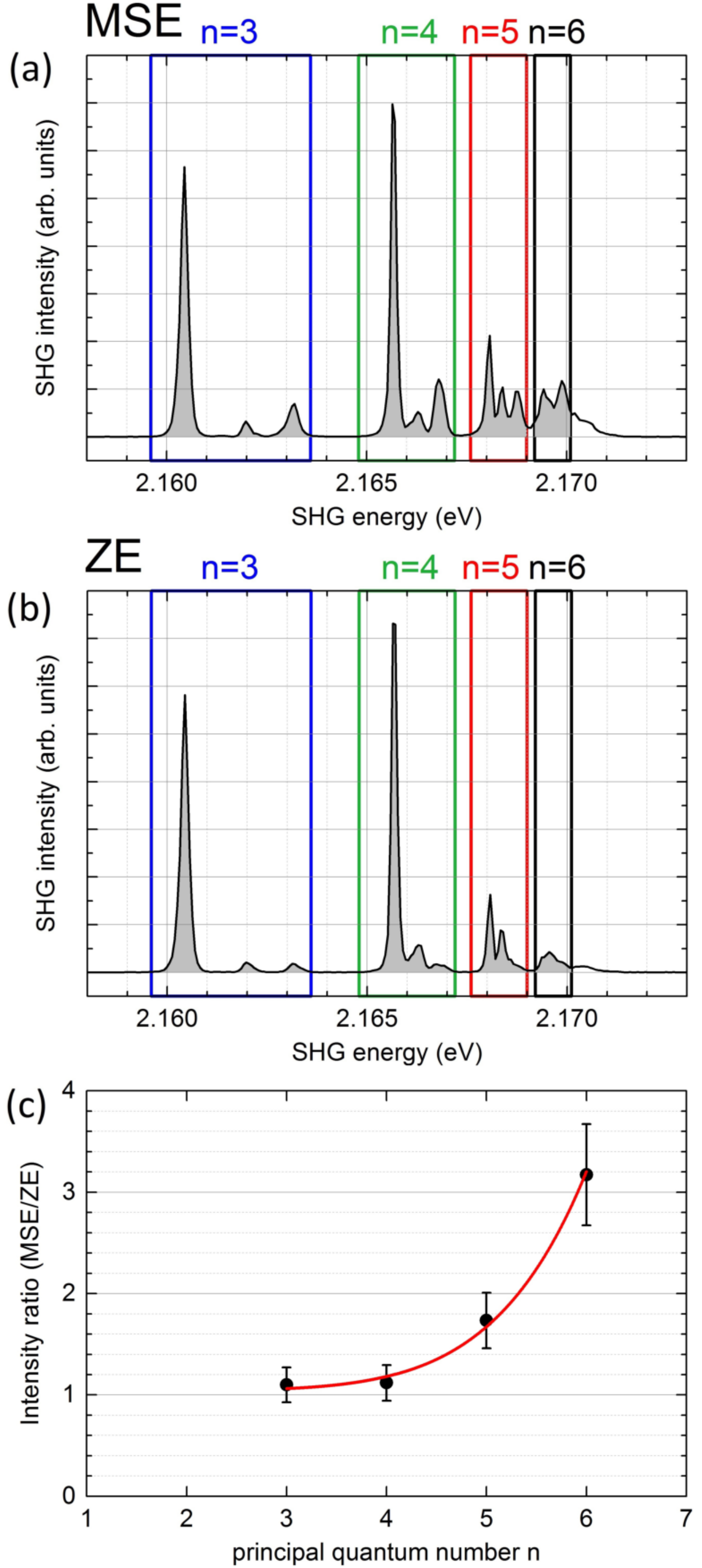}
		\caption[Scan]{
			Comparison of MSE (a) and ZE (b) induced SHG spectra for the different multiplets $n = 3, 4, 5, 6$ in a magnetic field of
			$B=\SI{1}{\tesla}$. The low field regime is chosen to minimize the spectral overlap of lines belonging to different multiplets as marked by the colored lines. The intensity within a multiplet is then determined by integration over all the lines within a multiplet. The ratio of the integrated SHG intensities induced by MSE relative to the one induced by ZE is plotted in (c) as function of the principal quantum number $n$ (full dots). The data are fitted by a power law function $R(n)=R_0+n^c$ with fit parameters $R_0 = 1.0 \pm 0.1$ and $c =6.4 \pm 1$.}
		\label{comparisonZEMSE}
	\end{center}
\end{figure}

The two upper panels in Fig.~\ref{comparisonZEMSE} show spectra exclusively induced by MSE (a) and ZE (b) covering the spectral range from $n= 3$ up to $n = 6$ at a magnetic field of $\SI{1}{\tesla}$, chosen to be low so that the field-induced splitting of the state multiplets belonging to different $n$ does not exceed the splittings between them. As before, the sample is oriented in such a way that $\textbf{k}\parallel[1\bar 1 0]$, $x\parallel\mathbf B\parallel[110]$ and $y\parallel[001]$ as shown in Fig.~\ref{fig:orientation}. The different multiplet ranges are marked by the differently colored boxes. One immediately sees that for the low lying excitons the SHG intensities show similar strength in both cases while for the $n = 5$ and $6$ multiplets the MSE induced SHG becomes dominant compared to the ZE-induced SHG. Moreover, for the low lying states, the SHG spectrum is dominated by one line with weak contributions from others, while for higher ones the SHG intensity is distributed over several lines, as might be expected from the larger state mixing due to the smaller energy separations between states within a multiplet corresponding to a certain principal quantum number $n$.

For a somewhat more quantitative analysis, we have integrated the SHG intensity recorded over the corresponding boxed energy range of a given $n$ and calculated the ratio of the SHG intensities of MSE relative to the one induced by the ZE. The result is shown by the circles in Fig.~\ref{comparisonZEMSE}(c) as a function of the principal quantum number $n$. One clearly sees an increase of the ratio with increasing $n$, starting from unity for $n=3$ and $4$, corresponding to equal MSE and ZE intensities. This behavior confirms the expectation from the microscopic theory that with increasing $n$ the MSE dominates over the ZE, see Eq.~\eqref{scaling}. By fitting the data with a power law function we obtain a scaling with power $6.4 \pm 1$, while the microscopic theory predicts a dependence scaling as $n^4$ for $n$ exceeding $\approx5$. The trend of a dominance of the MSE over the ZE is therefore consistent in experiment and theory, the deviation in the exponents may have different reasons one of which one is that the SHG spectra for $n \geqslant 4$ overlap already, so that state mixing becomes an important factor here.

\FloatBarrier

\section{Conclusions}
\label{sec.conclusion}

In this paper we present a comprehensive theoretical and experimental study of the nonlinear properties of excitons in Cu$_2$O in an external magnetic field. Two-photon absorption and second harmonic generation are considered. The focus is set on the forbidden crystalline directions, along which SHG is forbidden in the absence of magnetic field, so that only the magnetic-field-induced contributions arise. A detailed symmetry analysis gives us SHG polarization maps as functions of the linear polarization directions of the ingoing and outgoing waves. The polarization dependencies in the form of two-dimensional plots are very instructive to single-out the different SHG microscopic mechanisms: (i) the Zeeman effect related to the magnetic field-induced mixing and splitting of the exciton states of the same parity and (ii) the magneto-Stark effect resulting in the mixing of even- and odd-parity excitons by the combined action of the exciton motion and the magnetic field. Contributions of the various exciton states are identified by the symmetry analysis. The experimentally obtained SHG polarization dependences are in full agreement with the developed symmetry analysis, underlining its power in disclosing the nature of the optical transitions involved in SHG and TPA. We summarize in Appendix~\ref{appendix:polarization} calculated contour plots of the SHG intensity for different crystalline orientations and field configurations.

We have elaborated a microscopic theory of the nonlinear response on the excitons in Cu$_2$O and identified the pathways for the two-photon absorption and second harmonic generation. The microscopic theory explains why the $S$-mixed excitons of $\Gamma_5^+$ symmetry dominate the SHG induced by the Zeeman effect, while the $P$-excitons and the $D$ excitons of $\Gamma_1^+$ and $\Gamma_3^+$ symmetries provide much weaker contributions. The comparable contributions of $S$- and $P$-excitons to the SHG induced by the magneto-Stark effect are also explained. To simplify the analysis we have considered the mixing between different exciton states on the perturbative level, full non-perturbative calculations can be found in Ref.~\cite{RommelSHG}.

The developed theory and the experimental approaches can be readily extended for other materials with prominent exciton states. They can be also extended for searching other mechanisms, e. g. induced by external electric field or strain. 

The outstanding quality of the used Cu$_2$O crystals that is reflected by the extended series \cite{Brandt,Nature} of spectrally narrow exciton resonances (indicating high coherence) allows one to study the mechanisms of light-matter interaction in solids on an unprecedented level, as in most cases one can restrict to electric dipole transitions. Here, we have shown combinations of two dipole and a quadrupole transitions to explain the observed SHG. The coherence also allows identification of pronounced interference effects of interactions like demonstrated for the Zeeman and the magneto-Stark effect, leading to subtle state mixing effects between states of the same and different parity. This tunable mixing could allow, for example, excitation of particular exciton superposition states that can be uniquely identified through the two-dimensional plots of the SHG intensity as function of the linear polarizations of the fundamental wave and the SHG emission.

Our experimental setup allows pump-probe experiments with picosecond resolution. As an outlook we propose nonlinear optical experiments, where by time-resolved two-photon difference-frequency generation dynamical processes such as exciton-plasma or exciton-exciton interaction, see Ref.~\cite{plasma, Nature}, can be investigated. Further, second harmonic generation on paraexcitons, and in particular on the $1S$ paraexciton with an exceptionally narrow spectral line~\cite{Brandt}, is certainly another challenging spectroscopic task of interest. 

\begin{acknowledgments}
We acknowledge the Deutsche Forschungsgemeinschaft in the frame of the International Collaborative Research Centre TRR 160 (projects C8 and A8) and the Collaborative Research Centre TRR 142 (project B01). We also are grateful to the the Deutsche Forschungsgemeinschaft for the support by the project AS 459/1-3. M.M.G. and M.A.S. were partially supported by the Russian Foundation of Basic Research, project 19-52-12038. M.A.S. also acknowledges the financial support by the RFBR project 19-52-12063.
\end{acknowledgments}

\appendix

\section{Three band model for quadrupolar SHG}
\label{appendix:toy}

The results for the calculation of $\chi_c$ in Eq.~\eqref{SHG:cryst:111}, see also Eq.~\eqref{chi:c:micro:P}, are extremely cumbersome. That is why in this section for illustrative purposes we consider a three band model for SHG in a centrosymmetric crystal. We disregard the complex band structure and spin-orbit interaction and consider for simplicity conduction and valence bands of $S$-type ($\Gamma_1^+$ with Bloch amplitudes $\mathcal S_c$ and $\mathcal S_v$, respectively). We take into account intermediate states in the odd parity $\Gamma_4^-$ band composed of $\mathcal X$, $\mathcal Y$, $\mathcal Z$ Bloch states. We consider the incident radiation to be polarized along the $x$-axis which makes it possible to take into account as the intermediate state $s$ only the $\mathcal X$ state. Let $E_g'$ be the energy gap between the valence band and the $\Gamma_4^-$ remote band for which we assume that $E_g'\gg E_g$. We introduce the effective momentum matrix elements as
\[
P_{c,v} = \frac{\hbar}{m_0} \langle \mathcal X|\hat p_x|\mathcal S_{c,v}\rangle,
\]
and assume that $P_c$ and $P_v$ are real due to the choice of phases of basic functions.
Taking into account the $\mathbf k\cdot\mathbf p$ mixing of the bands in the lowest order we have for the two-photon transition matrix element
\begin{multline}
	\label{MTPA:toy}
	M^{(2)}(k_c,k_v) = \delta_{k_c,k_v+2q}\frac{P_cP_v}{E_g'}\\
	\times \left(1+ \frac{9 P_c^2+11 P_v^2}{(E_g')^2} k_v^2+\frac{10 P_c^2+8 P_v^2}{(E_g')^2} k_v q \right).
\end{multline}
Here $k_v$ and $k_c$ are the $x$-components of the electron wavevector in the initial (valence band) and final (conduction band) states, respectively. Similarly, the transition matrix element for a single photon emission from the conduction to the valence band state reads
\begin{equation}
\label{M1:toy}
M^{(1)}(k_v,k_c) = \delta_{k_v,k_c-2q} \frac{P_cP_v}{E_g'}(2q-2k_c).
\end{equation}

In order to calculate the susceptibility, the matrix elements $M^{(2)}(k_c,k_v)$ and $M^{(1)}(k_v,k_c)$ should be averaged over the exciton wavefunctions~\cite{pedersen,Glazov2017,semina:18}. For $S$-shell excitons the contribution to the susceptibility in leading order in $1/E_g'$ reads in agreement with Eq.~\eqref{chi:c:micro} of the main text
\begin{equation}
\label{chi:toy:S}
\chi \propto \left(\frac{P_cP_v}{E_g'}\right)^2 \frac{|\Phi_{nS}(0)|^2}{2\hbar\omega - E_g - E_{nS}+\mathrm i \Gamma_{nS}}.
\end{equation}
In this case the two-photon excitation of the $nS$ state is possible via dipole transitions and a quadrupolar process is needed for the exciton emission. In contrast, for $nP$ excitons the excitation is quadrupolar and requires and interference term $k_v q$ in Eq.~\eqref{MTPA:toy}. As a result [cf. Eq.~\eqref{chi:c:micro:P} of the main text]
\begin{equation}
\label{chi:toy:P}
\chi \propto \left(\frac{P_cP_v}{E_g'}\right)^2 \frac{|a_0\Phi_{nP}'(0)|^2}{2\hbar\omega - E_g - E_{nP}+\mathrm i \Gamma_{nP}},
\end{equation}
where
\[
a_0^2= \frac{A_c P_c^2+A_v P_v^2}{(E_g')^2}.
\]
Here $A_c$ and $A_v$ are the numerical coefficients determined by the electron and hole effective masses.
Note that $a_0$ has the dimension of length and is a combination of atomic scale parameters. That is why it is typically on the order of the lattice constant.

\section{SHG polarization diagrams for different crystal orientations}
\label{appendix:polarization}
\begin{figure*}[h]
	\begin{center}
		\includegraphics[width=0.99\textwidth]{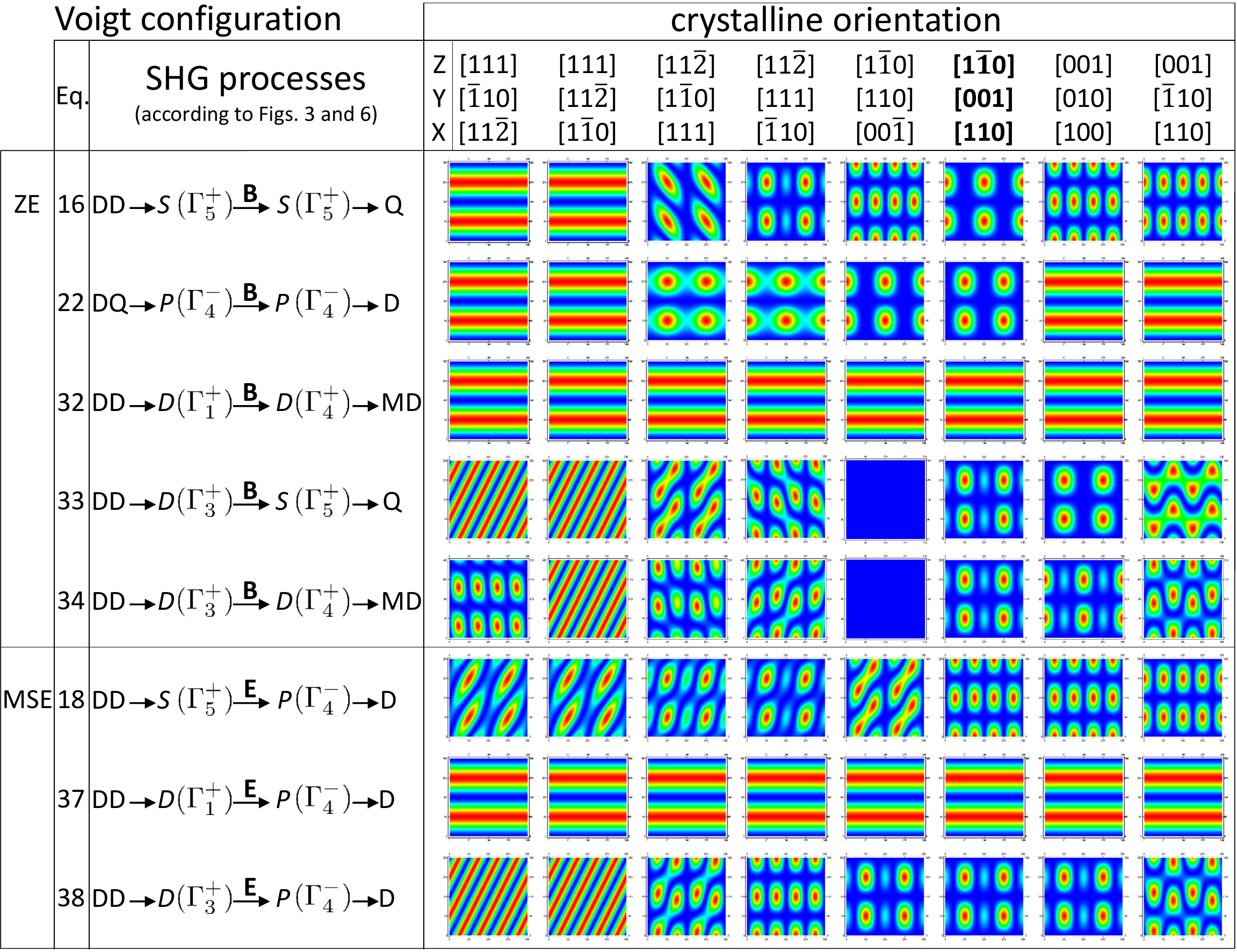}
		\caption[Scan]{$2$D polarization diagrams of SHG processes in Voigt configuration ($\mathbf k \parallel z$, $\mathbf E_{\text{MSE}} \parallel y$ and $\mathbf B \parallel x$) for selected crystalline orientations according to equations in Secs.~\ref{sec:dom} and \ref{sec:weak}. Experimental results are taken in the configuration ($z\parallel [1\bar10]$, $y\parallel [001]$ and $x\parallel [110]$) as marked by bold numbers. SHG processes correspond to the schematics of Figs.~\ref{fig:scheme} and \ref{fig:scheme_gamma31} with the following abbreviations: D - electric dipole transition, Q - electric quadrupole transition, MD - magnetic dipole transition, $S, P$ and $D$ - orbital quantum numbers, $\mathbf B$ - magnetic field, $\mathbf E$ - effective electric field.}
		\label{fig:voigt}
	\end{center}
\end{figure*}

\begin{figure*}[h]
	\begin{center}
		\includegraphics[width=0.99\textwidth]{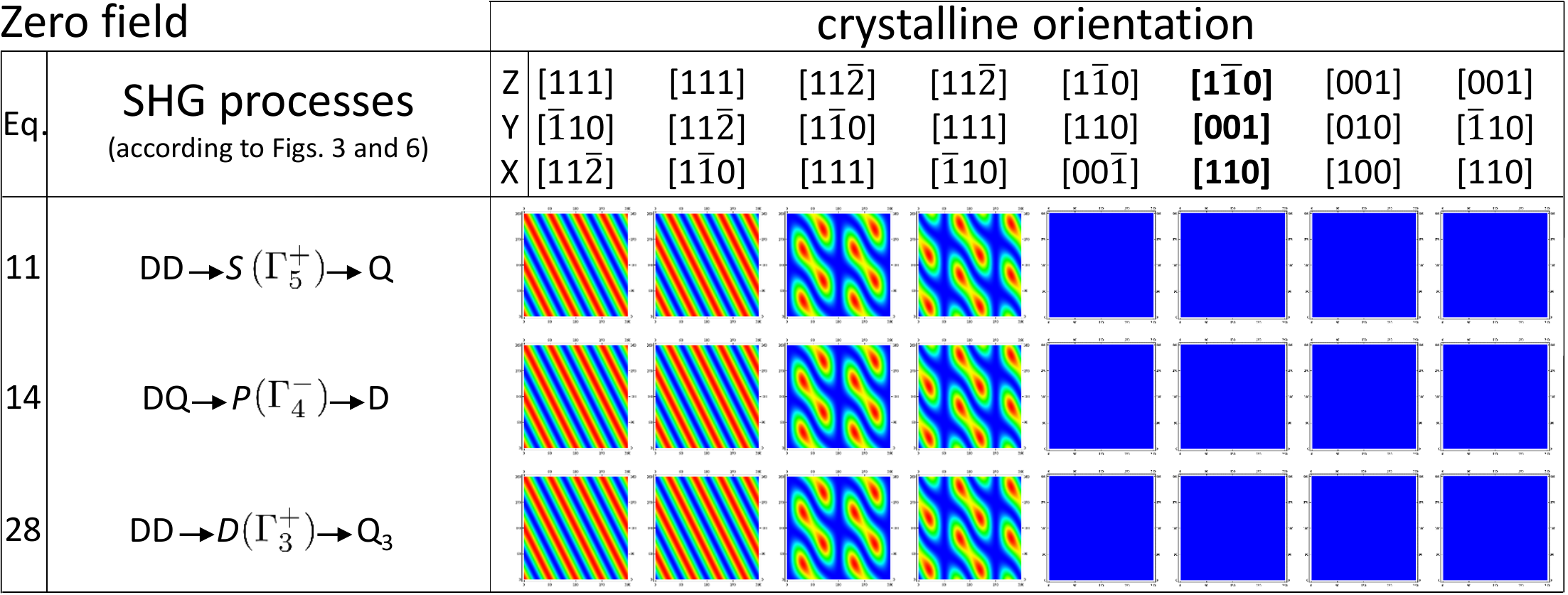}
		\caption[Scan]{$2$D polarization diagrams of zero-field SHG processes ($\mathbf k \parallel z$) for selected crystalline orientations according to equations in Secs.~\ref{sec:dom} and \ref{sec:weak}. SHG processes correspond to the schematics of Figs.~\ref{fig:scheme} and \ref{fig:scheme_gamma31} with the following abbreviations: D - electric dipole transition, Q and Q$_3$ - electric quadrupole transitions, $S, P$ and $D$ - orbital quantum numbers.}
		\label{fig:zero_field}
	\end{center}
\end{figure*}

\begin{figure*}[h]
	\begin{center}
		\includegraphics[width=0.99\textwidth]{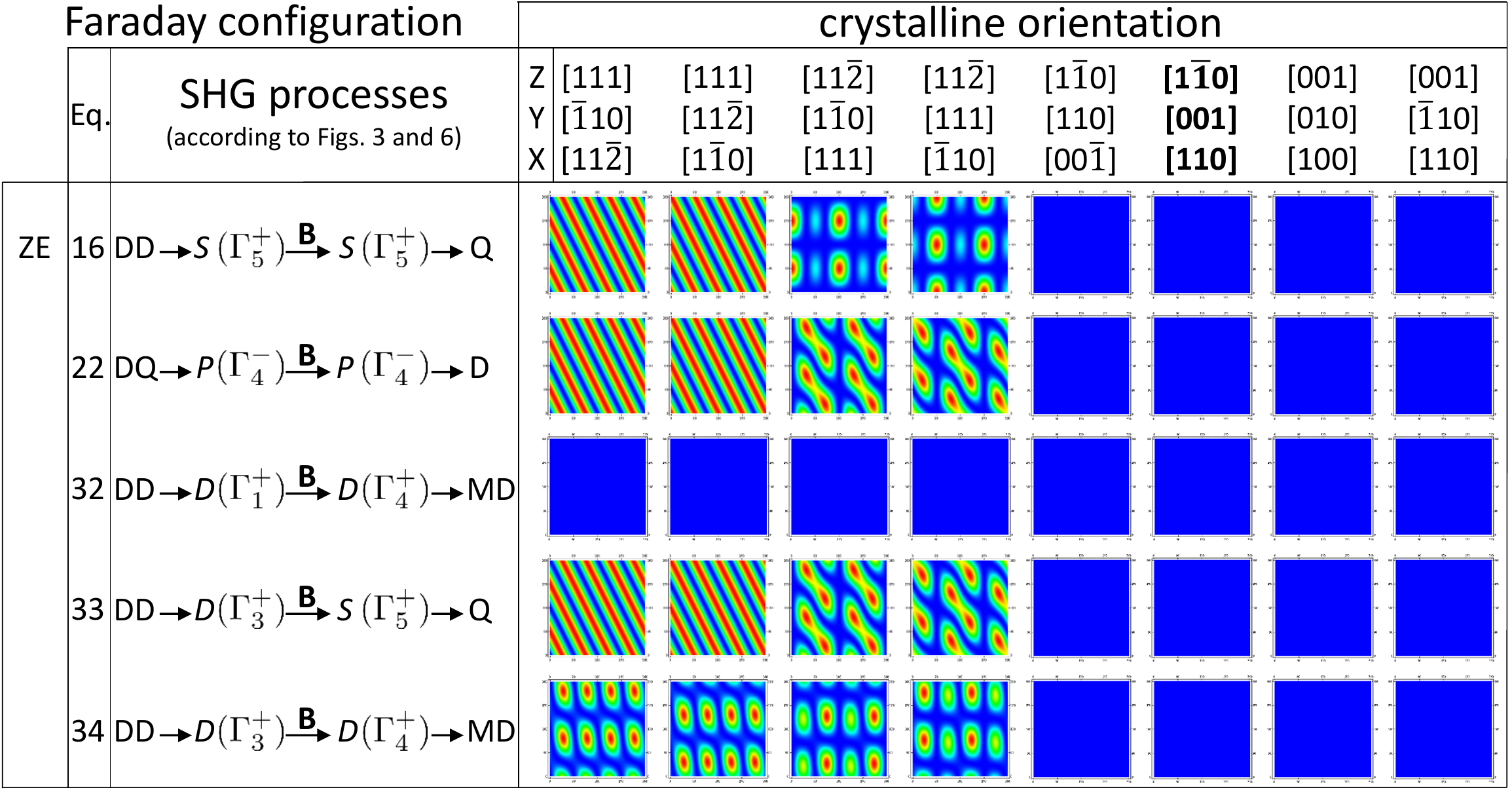}
		\caption[Scan]{$2$D polarization diagrams of Zeeman effect induced SHG processes in Faraday configuration ($\mathbf k \parallel \mathbf B \parallel z$) for selected crystalline orientations according to equations in Secs.~\ref{sec:dom} and \ref{sec:weak}. SHG processes correspond to the schematics of Figs.~\ref{fig:scheme} and \ref{fig:scheme_gamma31} with the following abbreviations: D - electric dipole transition, Q - electric quadrupole transition, MD - magnetic dipole transition, $S, P$ and $D$ - orbital quantum numbers, $\mathbf B$ - magnetic field.}
		\label{fig:faraday}
	\end{center}
\end{figure*}

In the manuscript we have presented theoretical and experimental $2$D polarization diagrams for selected crystalline and magnetic field orientations. We have distinguished \textit{Dominant Processes} [Sec.~\ref{sec:dom}, Fig.~\ref{fig:scheme}] and \textit{Weaker Processes} [Sec.~\ref{sec:weak}, Fig.~\ref{fig:scheme_gamma31}]. Our theoretical derivation of polarization dependences, however, applies for any crystalline and magnetic field orientation. In the following we extend the derivation for both processes to other crystalline orientations and present for each crystalline orientation $2$D polarization diagrams for two selected magnetic field orientations (for Faraday and Voigt configuration). As discussed in the main part of the manuscript [Secs.~\ref{sec:dom} and \ref{sec:weak}], the $2$D polarization diagrams [Fig.~\ref{fig:voigt}] allow to choose the appropriate crystalline and polarization configuration for the separation of different processes as e.g. Zeeman and magneto-Stark effect as well as dominant and weaker processes. As discussed in Ref.~\cite{Mund1} there is in the field-free case SHG only expected for $\mathbf k \parallel [111]$ and $[11\bar2]$ [Fig.~\ref{fig:zero_field}]. It turns out, that this applies also for SHG in Faraday configuration [Fig.~\ref{fig:faraday}]. It has to be noted, that SHG experiments with linearly polarized light in Faraday configuration have to take Faraday rotation in the sample as well as in cryostat windows into account.

\FloatBarrier

\end{document}